\documentclass[aps,prd,floats,preprint,nofootinbib]{revtex4} 
\usepackage{graphicx} 
\usepackage{amsmath} 
\usepackage{amsfonts} 
\usepackage{amssymb} 
 
\begin{document} 
\def\beq{\begin{equation}} 
\def\eeq{\end{equation}} 
\def\bea{\begin{eqnarray}} 
\def\eea{\end{eqnarray}} 
\def\ben{\begin{enumerate}} 
\def\een{\end{enumerate}} 
\def\ie{{\it i.e.}} 
\def\etc{{\it etc.}} 
\def\eg{{\it e.g.}} 
\def\lsim{\mathrel{\raise.3ex\hbox{$<$\kern-.75em\lower1ex\hbox{$\sim$}}}} 
\def\gsim{\mathrel{\raise.3ex\hbox{$>$\kern-.75em\lower1ex\hbox{$\sim$}}}} 
\def\ifmath#1{\relax\ifmmode #1\else $#1$\fi} 
\def\half{\ifmath{{\textstyle{1 \over 2}}}} 
\def\threehalf{\ifmath{{\textstyle{3 \over 2}}}} 
\def\quarter{\ifmath{{\textstyle{1 \over 4}}}} 
\def\eigth{\ifmath{{\textstyle{1\over 8}}}} 
\def\sixth{\ifmath{{\textstyle{1 \over 6}}}} 
\def\third{\ifmath{{\textstyle{1 \over 3}}}} 
\def\twothirds{{\textstyle{2 \over 3}}} 
\def\fivethirds{{\textstyle{5 \over 3}}} 
\def\fourth{\ifmath{{\textstyle{1\over 4}}}} 
\def\chitil{\wt\chi} 
\def\fbi{~{\mbox{fb}^{-1}}} 
\def\fb{~{\mbox{fb}}} 
\def\br{BR} 
\def\gev{~{\mbox{GeV}}} 
\def\calm{\mathcal{M}} 
\def\mll{m_{\ell^+\ell^-}} 
\def\tanb{\tan\beta} 
 
\def\wtil{\widetilde} 
\def\cnone{\wt\chi^0_1} 
\def\cnonestar{\wt\chi_1^{0\star}} 
\def\cntwo{\wt\chi^0_2} 
\def\cnthree{\wt\chi^0_3} 
\def\cnfour{\wt\chi^0_4} 
\def\snu{\wt\nu} 
\def\snul{\wt\nu_L} 
\def\msnul{m_{\snul}} 
\def\se{\wt e} 
\def\smu{\wt\mu} 
\def\snu{\wt\nu} 
\def\snul{\wt\nu_L} 
\def\msnul{m_{\snul}} 
 
\def\snue{\wt\nu_e} 
\def\snuel{\wt\nu_{e\,L}} 
\def\msnuel{m_{\snul}} 
 
\def\snubar{\ov{\snu}} 
\def\msnu{m_{\snu}} 
 
\def\snue{\wt\nu_e} 
\def\snuel{\wt\nu_{e\,L}} 
\def\msnuel{m_{\snul}} 
 
\def\snubar{\ov{\snu}} 
\def\msnu{m_{\snu}} 
\def\mcnone{m_{\cnone}} 
\def\mcntwo{m_{\cntwo}} 
\def\mcnthree{m_{\cnthree}} 
\def\mcnfour{m_{\cnfour}} 
\def\wt{\widetilde} 
\def\anti{\overline} 
\def\wh{\widehat} 
\def\cpone{\wt \chi^+_1} 
\def\cmone{\wt \chi^-_1} 
\def\cpmone{\wt \chi^{\pm}_1} 
\def\mcpone{m_{\cpone}} 
\def\mcpmone{m_{\cpmone}}

\def\staur{\wt \tau_R} 
\def\staul{\wt \tau_L} 
\def\stau{\wt \tau} 
\def\mstaur{m_{\staur}} 
\def\stauone{\wt \tau_1} 
\def\mstauone{m_{\stauone}} 
 
\def\gl{\wt g} 
\def\mgl{m_{\gl}} 
\def\stl{{\wt t_L}} 
\def\str{{\wt t_R}} 
\def\mstl{m_{\stl}} 
\def\mstr{m_{\str}} 
\def\sbl{{\wt b_L}} 
\def\sbr{{\wt b_R}} 
\def\msbl{m_{\sbl}} 
\def\msbr{m_{\sbr}} 
\def\sbot{\wt b} 
\def\msbot{m_{\sbot}} 
\def\sq{\wt q} 
\def\sqbar{\ov{\sq}} 
\def\msq{m_{\sq}} 
\def\slep{\wt \ell} 
\def\slepbar{\ov{\slep}} 
\def\mslep{m_{\slep}} 
\def\slepl{\wt \ell_L} 
\def\mslepl{m_{\slepl}} 
\def\slepr{\wt \ell_R} 
\def\mslepr{m_{\slepr}} 

\def\CC{{C\nolinebreak[4]\hspace{-.05em}\raisebox{.4ex}{\tiny\bf ++}}} 
\newcommand{ \slashchar }[1]{\setbox0=\hbox{$#1$}   
   \dimen0=\wd0                                     
   \setbox1=\hbox{/} \dimen1=\wd1                   
   \ifdim\dimen0>\dimen1                            
      \rlap{\hbox to \dimen0{\hfil/\hfil}}          
      #1                                            
   \else                                            
      \rlap{\hbox to \dimen1{\hfil$#1$\hfil}}       
      /                                             
   \fi}   
 
\vspace*{3cm} 
\title{Accurate Mass Determinations in Decay Chains with Missing 
  Energy: II} 
 
\author{Hsin-Chia Cheng${}^{a}$,  John F. Gunion${}^{a}$, Zhenyu Han${}^{a}$,  and Bob 
  McElrath${}^{b}$} 
 
\affiliation{ \small \sl ${}^{a}$Department of Physics, University of 
  California, Davis, CA 95616\\  
${}^{b}$CERN, Geneva 23, Switzerland}

\begin{abstract}  
  We discuss kinematic methods for determining the masses of 
  the particles in events at a hadron collider in which a pair of 
  identical particles is produced with each decaying via a series of 
  on-shell intermediate beyond-the-SM (BSM) particles to visible SM 
  particles and an invisible particle (schematically, $pp\to ZZ+jets$ 
  with $Z\to Aa \to Bba \to Ccba\to \ldots\to cba\ldots+ N$ where 
  $a,b,c,\ldots$ are visible SM particles or groups of SM particles, 
  $A,B,C,\ldots$ are on-shell BSM particles and $N$ is invisible). 
  This topology arises in many models including SUSY processes such as 
  squark and gluino pair production and decay. We present the detailed 
  procedure for the case of $Z\to 3~\mbox{visible particles}+N$ and 
  demonstrate that the masses obtained from the kinematic procedure 
  are independent of the model by comparing SUSY to UED.   
\end{abstract} 
 
\maketitle 

\section{Introduction} 
\label{sec:introduction} 
 
Many solutions to the hierarchy problem require new particles whose 
loop corrections to the Higgs mass-squared cancel the quadratically 
divergent Standard Model (SM) loop corrections. The masses of the new 
particles (especially the particle or particles which cancel the SM 
top loop) should be sufficiently small that the Higgs mass can be 
naturally  below the TeV scale. On the other hand, if the new 
particles have masses below $\mathcal{O}({\rm TeV})$ then LEP 
observables will be strongly affected if they are exchanged at tree 
level or can be singly produced. Both these latter possibilities are 
automatically removed if there is a symmetry under which the new 
particles are odd and the SM particles are even. In particular, the 
new particles can then only contribute to the electroweak observables 
at the loop level, and new particles with masses of order a few 
hundreds of GeV can be compatible with the data. In such scenarios the 
lightest of the new particles is automatically stable and it should be 
neutral for consistency with bounds on new charged stable matter. 
Typically, it is also weakly interacting.  Such a weakly interaction 
massive stable particle (WIMP) will be invisible, leading to ``missing 
energy'' in particle detectors. This scenario is also highly desirable 
since such a WIMP can readily provide the dark matter known to be 
present in the universe. 
 
Almost all the models with dark matter candidates also contain 
additional particles that are charged not only under the new symmetry 
but also carry SM ``charges,'' most often  including color. At a 
collider, these new particles must (and will)  be pair-produced, and 
since they are heavier than the dark matter particle, they will 
cascade decay down to it.  In many cases, this cascade radiates SM 
particles in a series of $A\to Bc$, $1-body \to 2-body$ decays, in 
which $A$ and $B$ are new physics particles while $c$ is a SM 
particle.  (In some cases, phase space restrictions force one of the 
new particles off-shell and $A\to B^* c\to C d c$, $1-body\to 3-body$ 
decays are relevant.) Since the final step in the chain will yield a 
dark matter particle, the typical collider signals for such a scenario 
will be jets and/or leptons plus missing energy.

Supersymmetry (SUSY) is the most popular model of this type.  In SUSY, 
the new symmetry is termed matter Parity (sometimes called 
$R$-parity). Its conservation implies that the Lightest Supersymmetric 
Particle (LSP) is stable. In most supersymmetric models the LSP is the 
lightest neutralino, which is a good dark matter candidate. It appears 
at the end of every supersymmetric particle decay chain and escapes 
the detector. All supersymmetric particles are produced in pairs, 
resulting in at least two missing particles in each event. 
 
Other theories of TeV scale physics with dark matter candidates have 
been recently proposed. They have experimental signatures very similar 
to SUSY: {\it i.e.} multi leptons and/or jets plus missing energy. For 
instance, Universal Extra Dimensions (UEDs)~\cite{ued,Cheng:2001an}, 
little Higgs theories with $T$-parity (LHT)~\cite{lht}, and warped 
extra dimensions with a $Z_3$ parity~\cite{Agashe:2004ci} belong to 
this category of models.   
 
Clearly, being able to reconstruct events with missing energy is an 
important first step to distinguish various scenarios and establish 
the underlying theory.  In addition, studies~\cite{Baltz:2006fm} 
suggest that the mass of the dark matter particle, and the masses of 
any other particles with which it can coannihilate, need to be 
determined to within a few GeV in order to be able to compute the dark 
matter density in the context of a given model. A very important 
question is then whether or not the LHC can achieve such accuracy or 
will it be necessary to wait for threshold scan data from the ILC. 
The goal of this paper will be to provide details regarding the 
kinematic techniques developed in 
Refs.~\cite{Cheng:2007xv,Cheng:2008mg} that provide the needed 
accuracy using just LHC data. For the case of 3 visible particles per 
decay chain, the focus of this paper, we also show that the kinematic 
technique gives masses for the BSM particles that are completely 
insensitive to the particular model by comparing a SUSY case to a UED 
case where the decaying BSM and final invisible BSM particles in the 
two cases have the same masses.  This implies that it is unnecessary 
to determine the overall mass scale of the BSM particles using  
model-dependent information, such as total cross sections. Indeed, to 
fully test a potential model, it is {\it necessary} to first determine 
the masses of the produced particles just based on kinematic 
information.  Once the masses are known, there are many chain decay 
configurations for which it will be possible to determine the 
four-momenta of all the particles on an event-by-event basis. The 
four-momenta can then be employed in computing the matrix element 
squared for different possible spin assignments.  In this way, a spin 
determination may be possible which, in combination with cross section 
information, can be used to distinguish different models.

In recent years there have been numerous studies in the context of 
SUSY-like theories of how to measure the super-partner masses just 
based on kinematic information~\cite{Hinchliffe:1996iu, mt2, 
  Bachacou:1999zb, Allanach:2000kt,Gjelsten:2004ki, Kawagoe:2004rz, 
  Lester:2005je, Arkani-Hamed:2005px,Butterworth:2007ke,Lester:2007fq, 
  kink, Barr:2007hy, m2c, hybrid,  hiddenthreshold, Cheng:2008hk, 
  Burns:2008va, Konar:2008ei, Papaefstathiou:2009hp, Burns:2009zi}.  
In some cases the procedures employ a single long decay chain of 
super-particles, usually requiring 3 or more visible particles in the 
decay chain in order to have enough invariant mass combinations of the 
visible particles to achieve sensitivity to the absolute mass scale, 
as opposed to simply mass differences.  Squark decay, 
Fig.~\ref{fig:chain_decay}, is an example of one such chain. 
\begin{figure} 
\begin{center} 
\includegraphics[scale=0.6,angle=0]{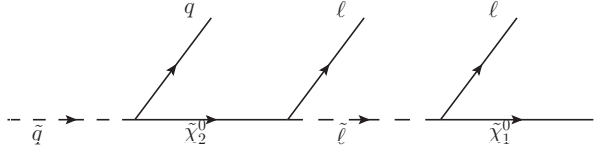} 
\end{center} 
\caption{\label{fig:chain_decay} 
 A decay chain in SUSY.} 
\end{figure} 

Our approach has been to pursue alternative procedures that employ 
events in which a pair of identical (particle and antiparticle, {\it 
  e.g.} squark plus anti-squark) BSM particles is produced and both 
decay in the same manner.  In such an event, information from both 
decay chains in the event can be included at once. In our first paper 
\cite{Cheng:2007xv}, we tackled the difficult case where we assumed 
that only two particles appeared in each chain decay, {\it e.g.} 
making use only of the leptons appearing in 
Fig.~\ref{fig:chain_decay}.  In this case, kinematic constraints  
alone can not give a discrete solution for the 
unknown masses.  Nonetheless, the space of the allowed solutions does 
contain enough information about the new particle masses and they can be  
extracted using a statistical procedure~\cite{Cheng:2007xv}. The mass 
determination can be further improved by combining with other kinematic 
variables such as $M_{T2}$~\cite{mt2,Cheng:2008hk}. However, these 
kind of analyzes usually require large statistics in order to achieve a 
reasonable precision. In this paper, we provide details on the case 
where 3 visible particles are present in each chain decay.  In this 
case, a single pair of events provides enough information to yield a 
discrete set of possible masses. Therefore, very few events are needed 
for a rough determination of the masses, in contrast to the 
statistical methods which rely on the availability of a large number 
of events.

The general topology on which we focus is then that of 
Fig.~\ref{fig:topology}.   
\begin{figure} 
\begin{center} 
\includegraphics[scale=0.4,angle=0]{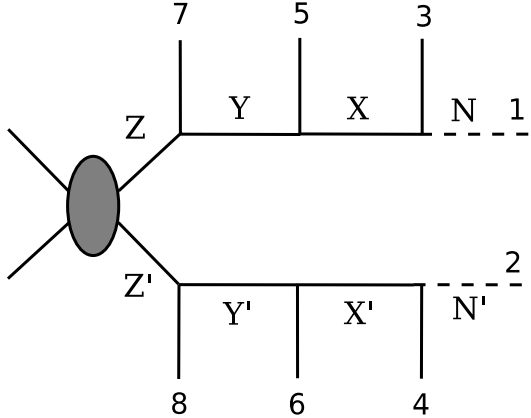} 
\end{center} 
\caption{\label{fig:topology} 
 The event topology we consider.} 
\end{figure} 
After including combinatorics and resolution, 
we will achieve root-mean-square (rms) accuracies on the three 
underlying masses in the decay chain of order a few GeV (depending 
upon the number of available events) with a  systematic shift that can 
be easily corrected for.  This result is fairly stable when 
backgrounds are included so long as $S/B\gsim 2$. 
 
The organization of the paper is as follows.  In 
Sec.~\ref{sec:counting}, we give the general counting of constraints 
and unknowns for single chain and multiple chain events.  In 
Sec.~\ref{sec:topology}, we give a more detailed exposition regarding 
solving the topology of Fig.~\ref{fig:topology}.  In 
Sec.~\ref{sec:applications}, we first demonstrate how the masses of 
the $Z$, $Y$, $X$ and $N$ particles in Fig.~\ref{fig:topology} can be 
very precisely determined using just a few events if there are no 
effects associated with combinatorics, particle momentum measurement 
resolutions or backgrounds. We then develop the very crucial 
strategies for dealing with the realistic situation where 
combinatorics, resolution effects and backgrounds are present.  We 
still find good accuracies for all the masses using only the kinematic 
information contained in the available events. We study the accuracy 
of the mass determinations as a function of the available number of 
events and as a function of the signal to background ratio. In 
sec.~\ref{sec:UED}, we compare results for the SUSY and UED cases and 
show that the masses determined are independent (to within one to two 
GeV) of which model is 
employed.  We summarize and present additional discussion in 
Sec.~\ref{sec:discussion}. Some of the material in 
sec.~\ref{sec:topology} and sec.~\ref{sec:applications} has appeared 
in Ref.~\cite{Cheng:2008mg}, but is included in the present article 
for completeness and to simplify some of the discussions.   
 
\section{Constraints Counting} 
\label{sec:counting} 
 
To begin, it is useful to perform a general counting of 
observables and constraints for various different configurations. 
We consider first the counting when only one decay chain in the event 
is considered at a time. We then show the increase in constraints 
possible if both decay chains in each event are considered at once. 
 
\subsection{Single decay chain case}

We begin with the chain decay $X\to a Y \to a b N$, where the 
4-momenta of the SM-particles $a$ and $b$ are directly measured. For 
each event, there are four unknowns due to the unobserved 4-momentum 
of the $N$. In addition, we have the three unknown masses (the 
same for every event) $m_X$, $m_Y$, and $m_N$. These are subject to $3n$ 
constraints coming from requiring that the $X$, $Y$, and $N$ be on 
their mass-shell.  Thus, after $n$ events we have $3+4n-3n=3+n$ free 
parameters.  No matter how many events we examine, we will not be able 
to obtain a discrete solution (or set of solutions) for $m_X$, $m_Y$, 
and $m_N$. 
 
Next, consider a decay chain with three observable SM 
particles: $Z\to a Y \to a b X \to  a b c N $.  In this case, the 
number of unknowns after $n$ events is $4+4n$ and the number of 
constraints is $4n$ (the $Z$, $Y$, $X$ and $N$ masses, which are the 
same for every event).  After $n$ events we then have $4+4n-4n=4$ free 
parameters, which basically correspond to the 4 unknown masses of the 
decaying particles. In this case, each event will determine a region 
in the $(m_Z,m_Y,m_X,m_N)$ mass space and, as more events are 
accumulated, in an ideal world the region of mass space consistent 
with all events would become more and more restricted, but it will 
never reach a discrete point (or a set of discrete points) with any 
number of events. To pin down the actual mass point, one needs to use 
additional information by examining the end points of certain 
kinematic distributions. 
 
If one considers (as in \cite{Kawagoe:2004rz}) a chain with four fully 
measured SM-particles, $A \to aZ \to abY \to abcX \to abcdN$, then  
we have 5 on-shell particles, and after $n$ events, we end up with 
$5+4n$ unknowns and $5n$ constraints.  
The number of free parameters is then $5-n$, implying that (up to 
discrete ambiguities associated with a high order polynomial and 
ignoring combinatorics and resolution) $n=5$ events would be 
sufficient to solve for the resonance masses. However, combinatorics 
and resolution effects will considerably complicate the situation, as 
already apparent from the study of \cite{Kawagoe:2004rz}, where they 
assume that $m_Y$, $m_X$ and $m_N$ are known, leaving, in principle, $2+4n$ 
unknowns and $5n$ constraints after $n$ events, implying that only 
$n=2$ events would be needed to solve for the remaining $2$ masses, 
$m_A$ and $m_Z$. After including combinatorics and resolution, 
Ref.~\cite{Kawagoe:2004rz} needed many more than 2 events in order to 
get a mass determination. 
 
The general counting procedure is apparent.  If the decay chain has 
$N_A-1$ on-shell decaying particles and a final invisible particle, 
then after $n$ events there will be $4n$ unknowns associated with the 
4-momenta of the invisible particle.  There will also be $N_A$ 
unknowns corresponding to the unknown masses of all the BSM 
particles. These unknowns will be subject to $N_An$ constraints from 
the requirement that all the BSM particles have the same on-shell 
masses in each event.  The number of unknowns after $n$ signal events 
will then be  
\beq 
N_U=N_A+4n-N_An=N_A+(4-N_A)n\,. 
\eeq 
For $N_A\leq 4$, a discrete solution or set of solutions is not 
possible regardless of how many events are available. The actual masses  
may still be obtained by combining additional information from the 
kinematic distributions for $N_A=4$ (\eg\ the squark decay case of 
Fig.~\ref{fig:chain_decay}). For $N_A=5$ (\eg\ the SUSY $\wtil g\to 
q\wtil q\to qq\cntwo\to qql\slep\to qqll\cnone$ decay chain) $n=5$ 
events will give a discrete set of solutions for the masses.  Still 
longer decay chains would require fewer events to obtain a set of 
discrete possibilities for the BSM particle masses.  An example of a 
longer decay chain with $N_A=6$ (implying that 3 events would give a 
discrete set of solutions for the masses) would be  
\beq 
\wtil g\to q\wtil q\to qq\cnthree\to qqZ 
\cntwo\to qqZl\slep\to qqZll\cnone\,, 
\label{longchain} 
\eeq 
where the $Z$ would be observed 
in one of its visible decay modes. By considering the full event at 
once through inclusion of information from both decay chains, one need 
not resort to such long decay chains in order to get to the point of 
having a discrete set of solutions for the masses.

\subsection{Using the whole event, \ie\ both decay chains} 
 
When considering the whole event at once, the constraint counting 
proceeds differently. Assuming that there are two invisible particles 
present in the final state, the number of unknowns associated with 
their 4-momenta in all $n$ events is $8n$.  Requiring that the sum of the 
invisible transverse momenta equal minus the sum of the visible 
transverse momenta imposes $2n$ constraints on these unknowns.  In 
addition, let us suppose that the topology is such that $N_B$ masses are 
unknown ($N_B$ includes the unknown masses of the invisible particles, 
which we do not require to be the same at this point in our counting). 
Each event will also be subject to a number $N_A$ of on-shell mass 
constraints, including the requirement that the two invisible 
particles have masses equal to their on-shell (unknown) values.  Then, 
after $n$ events there are $N_An$ constraints.  Thus, the number of 
unknowns before imposing constraints is $N_B+6n$ and the number of 
constraints is $N_An$, leaving 
\beq 
N_U=N_B+(6-N_A)n 
\eeq 
unknowns. If we consider an event  
with $N_A -2>4$ on-shell decays and require that there be no unknowns, {\it 
  i.e.}  $N_U \leq 0$, after $n_S$ events, we find 
\beq 
n_S\geq {N_B\over N_A-6}\,. 
\eeq   
Of course, in general $N_B\leq N_A$.  For symmetric chains, 
$N_B=N_A/2$. For chains in which only the final missing particles are 
assumed to have the same mass, $N_B=N_A-1$.  The particular case we 
focus on in this paper, Fig.~\ref{fig:topology}, corresponds $N_A=8$ 
and $N_B=4$. In this case, $n_S=2$ events will lead to $N_U=0$, 
implying a discrete set of solutions for the unknown masses of 
$Z,Y,X,N$. Were we to consider the case of two identical chains with 
only 2 visible particles in each and unknown masses for $Y,X,N$, one 
would have $N_A=6$ and $N_B=3$, leading to $N_U=3$ (corresponding to 
the unknown $Y,X,N$ masses). This is the case considered in 
\cite{Cheng:2007xv};  a discrete 
set of solutions will never emerge with any finite number of signal events.  
Additional information from kinematic distributions is needed to 
pin down the masses. Conversely, if one goes to symmetric longer decay 
chains (such as that of Eq.~(\ref{longchain})) with $N_A=12$ and 
$N_B=6$, then just one event will be sufficient to give a discrete set 
of mass solutions. The general problem with longer decay chains is that 
they are less likely to occur and harder to identify even if they occur, 
as the visible particles may be lost or hard to isolate.  
The shortest decay chains that can give rise to discrete solutions, \ie,  
the topology of Fig.~\ref{fig:topology}, 
with $N_A=8$ and $N_B=4$ yielding $n_S=2$, is likely to be the  
optimal configuration for the mass determination.  
 
Extension to more missing particles, or quadratic constraints is 
straightforward, but increases the order of the equations to be 
solved, requiring more advanced polynomial solvers.  It also requires 
that more masses be specified. For example, let us assume some 
definite topology for which each event contains 3 `missing' particles. 
If we do not presume any mass equalities among the total number, $N_A$, 
of decaying resonances and missing particles, then the number of 
unknowns after $n$ events is $N_A+(4\times 3-2)n$, where the $-2$ is the 
transverse momentum constraint setting the visible transverse momentum 
equal to minus the invisible transverse momentum in each event and the 
$4\times 3=12$ just corresponds to the 4 unknown components of each of 
the 3 invisible particles' 4-momenta.  The number of mass-shell 
constraints is (by definition) $N_An$. If the topology is such that
$N_A>10$ on-shell masses can be reconstructed from the visible momenta
and the invisible momenta and if only $N_B$ of the $N_A$ masses are
independent, then the number of unknowns after $n$ events is
$N_B+(4\times 3-2)n$ and the number of constraints is $N_An$. Again
neglecting possible relations among these masses and requiring the
final number of unknowns, $N_U=N_B+10n - N_An$, to be 0 or negative in
order to have fewer unknowns than constraints after $n=n_S$ events,
one finds that 
\beq  
n_S \geq{N_B\over (N_A-10)}  
\eeq  
events would lead to a certain set of discrete mass solutions. If all
the invisible particles are the same then $N_B\leq N_A-2$.
 
We have classified many possible decay chains which fall into a category 
such that a small handful of events could give an essentially unique 
mass spectrum in the absence of combinatorial and experimental 
resolution effects.  However, generically speaking one wishes to keep 
the chains as short as possible while consistent with a small number of 
events being sufficient to yield a discrete spectrum of mass solutions. 
This is because (a)~shorter chains are easier to isolate on an 
event-by-event basis and (b)~combinatorial and resolution smearing of 
the solutions may be lessened.

\section{Basic equations for the topology of Fig.~2} 
\label{sec:topology} 
 
The topology on which we focus in this paper is that given in 
Fig.~\ref{fig:topology}. As sketched in the previous section,  
this is an ideal topology for precise mass 
reconstruction. 
 Assuming $m_N=m_{N'}$, $m_X=m_{X'}$, 
$m_Y=m_{Y'}$, $m_Z=m_{Z'}$ and denoting the 4-momenta for particles $i 
(i=1\ldots 8)$ with $p_i$, we have 
\begin{eqnarray} 
&&p_1^2=p_2^2(=m_N^2),\label{mass_shell}\\ 
&&(p_1+p_3)^2=(p_2+p_4)^2(=m_X^2),\\ 
&&(p_1+p_3+p_5)^2=(p_2+p_4+p_6)^2(=m_Y^2),\\ 
&&(p_1+p_3+p_5+p_7)^2=(p_2+p_4+p_6+p_8)^2(=m_Z^2). 
\end{eqnarray} 
We assume further that the only invisible particles are particles $1$ 
and $2$, and thus have two more constraints, 
\begin{eqnarray} 
p_1^x+p_2^x&=&p_{miss}^x,\quad 
p_1^y+p_2^y=p_{miss}^y.\label{misspt} 
\end{eqnarray} 
There are 8 unknowns in Eqs.~(\ref{mass_shell}) through 
(\ref{misspt}), namely, the 4-momenta $p_1$ and $p_2$ of the missing 
particles. Therefore the system is underconstrained and we cannot 
solve the equations. This situation changes if we add a second event 
with the same decay chains.  Denoting the 4-momenta in the second 
events as $q_i$ $(i=1\ldots 8)$, we have 8 more unknowns, $q_1$ and 
$q_2$, but 10 more equations, 
\begin{eqnarray} 
&&q_1^2=q_2^2=p_1^2,\\ 
&&(q_1+q_3)^2=(q_2+q_4)^2=(p_2+p_4)^2,\\ 
&&(q_1+q_3+q_5)^2=(q_2+q_4+q_6)^2=(p_2+p_4+p_6)^2,\\ 
&&(q_1+q_3+q_5+q_7)^2=(q_2+q_4+q_6+q_8)^2=(p_2+p_4+p_6+p_8)^2,\\ 
&&q_1^x+q_2^x=q_{miss}^x,\quad 
q_1^y+q_2^y=q_{miss}^y.\label{misspt2} 
\end{eqnarray}  
Altogether, we have 16 unknowns and 16 equations. The system can be 
solved numerically and we obtain discrete solutions for $p_1$, $p_2$, 
$q_1$, and $q_2$ and thus the masses $m_N$, $m_X$, $m_Y$, $m_Z$. Note that 
the equations always have 8 complex solutions, but we will keep only 
the real and positive-energy ones which we simply call ``solutions'' in the 
rest of the paper. Thus, up to a certain number of discrete 
ambiguities we can determine the $Z,Y,X,N$ masses by pairing any two 
signal events. Even a few pairs of events are typically sufficient to 
eliminate the discrete ambiguities due to higher order equations.   
However, effects such as wrong 
combinations and solutions, initial and final state radiation, 
experimental resolutions, and background events will add 
complications, which we address in Sec.~\ref{sec:applications}.

 
The equations (\ref{mass_shell}) through (\ref{misspt2}) can be easily 
reduced to 3 quadratic equations plus 13 linear equations, 
\begin{eqnarray} 
&&p_1^2=p_2^2=q_1^2=q_2^2,\label{eq:constraint1}\\ 
&&2p_1\cdot p_3+p_3^2=2p_2\cdot p_4+p_4^2=2q_1\cdot q_3+q_3^2=2q_2\cdot q_4+q_4^2,\\ 
&&2(p_1+p_3)\cdot p_5 +p_5^2=2(p_2+p_4)\cdot p_6+p_6^2=\nonumber\\ 
&&\    \ =2(q_1+q_3)\cdot q_5+q_5^2=2(q_2+q_4)\cdot q_6+q_6^2,\\ 
&&2(p_1+p_3+p_5)\cdot p_7 +p_7^2=2(p_2+p_4+p_6)\cdot p_8+p_8^2=\nonumber\\ 
&&\    \ =2(q_1+q_3+q_5)\cdot q_7+q_7^2=2(q_2+q_4+q_6)\cdot q_8+q_8^2,\\ 
&&p_1^x+p_2^x=p_{miss}^x,\  \ p_1^y+p_2^y=p_{miss}^y,\\ 
&&q_1^x+q_2^x=q_{miss}^x,\  \ q_1^y+q_2^y=q_{miss}^y\,, \label{eq:constraint6}
\end{eqnarray}  
where all but the first line are linear equations because 
$p_{3,4,5,6,7,8}$ and $q_{3,4,5,6,7,8}$ are all visible measured momenta. 
In general, the above equation system has 8 complex solutions, each of 
which could be real. This can be shown by calculating the Gr\"obner 
basis \cite{groebner}, in which the system is transformed to an 8th 
order univariate equation plus 15 linear equations.  Since the other 
15 equations are linear, it is straightforward to solve for the other 
15 variables once the 8th order equation is solved. Commercial 
software such as Mathematica uses this method. However, it consumes an 
intolerably long time for a single or small number of PCs. We take a simpler 
and faster approach which is described in detail in the Appendices.  
In our method, instead of ending up with an 8th order equation, we 
obtain a 9th order univariate polynomial equation and therefore 
introduce a fake solution in addition to the true solutions.   
The 9th order univariate polynomial equation is 
 numerically solved using algorithm TOMS/493 
\cite{toms493}. The fake solution can be easily eliminated  
by substituting back all solutions in 
the original equations.

\section{Applications}\label{sec:applications} 
\subsection{SUSY point SPS1a} 
For illustration and easy comparison to the literature, we apply our 
method for the SUSY point, SPS1a \cite{Allanach:2002nj},  
although many of the discussions below apply for generic cases. For 
SPS1a, the particles corresponding to $N,X,Y,Z$ are $\cnone$, 
$\slep_R(\ell=e/\mu)$, $\cntwo$, $\sq_L(q=d,u,s,c)$ 
respectively.  The masses are  
\bea  
&&m_N=97.4\gev,\quad 
m_X=142.5\gev,\cr  
&&m_Y=180.3\gev,\quad m_Z=564.8/570.8\gev\,, 
\label{minsps1a} 
\eea  
with the final two numbers corresponding to up/down type squarks 
respectively.  Since $m_{\wtil \tau}\neq m_{\wtil e,\wtil\mu}$, the 
$\ell=\tau$ case is an important background. We generate events with 
PYTHIA 6.4 \cite{Sjostrand:2006za}.

We first consider the ideal case: no background events, all visible 
momenta measured exactly, all intermediate particles on-shell and each 
visible particle associated with the correct decay chain and position 
in the decay chain. We also restrict the squarks to be up-type only. 
In this case, we can solve for the masses exactly by pairing any two 
events. The only complication comes from there being 8 complex 
solutions for the system of equations, of which more than one can be 
real and positive. Of course, the wrong solutions are different from 
pair to pair, but the correct solution is common. The mass 
distributions for the ideal case with 100 events (no kinematic cuts 
applied) are shown in Fig.~\ref{fig:ideal}. Note the logarithmic 
scale. As expected, we observe $\delta$-function-like mass peaks on 
top of small backgrounds coming from wrong solutions. On average, 
there are about 2 solutions per pair of events. 
 
\begin{figure} 
\begin{center} 
 \includegraphics[width=0.6\textwidth]{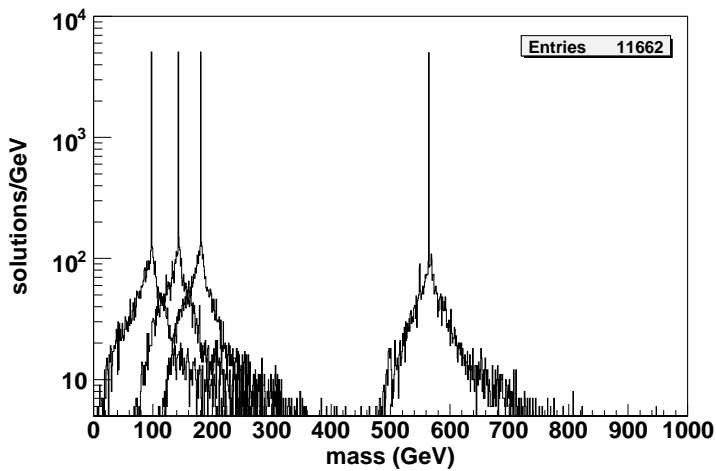} 
\vspace*{-.1in} 
\caption{\label{fig:ideal}We plot the number of mass solutions (in 1 
  GeV bins --- the same binning is used for the other plots) vs. mass 
  in the ideal case. All possible pairs for 100 events are 
  included. Signal events only.} 
\vspace*{-.3in} 
\end{center} 
\end{figure} 
The $\delta$-functions in the mass distributions arise only when exactly 
correct momenta are input into the equations we solve.   
To be experimentally realistic, we now include the following. 
\begin{figure} 
\begin{center} 
 \includegraphics[width=0.6\textwidth]{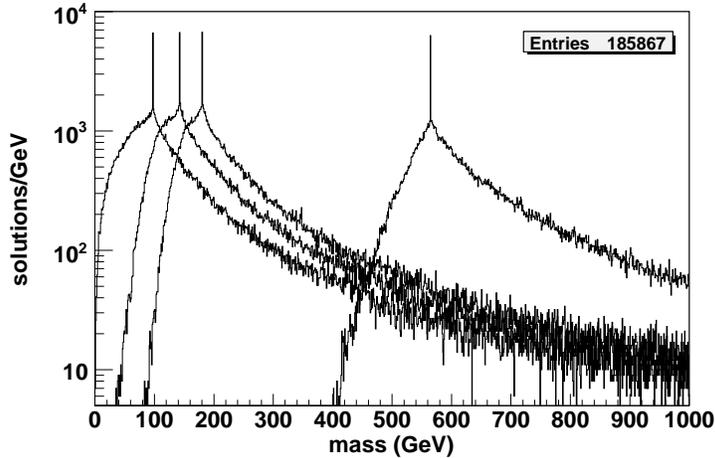} 
\vspace*{-.1in} 
\caption{\label{fig:allcombis}Number of mass solutions versus mass after 
  including all combination pairings for 100 events. Signal events 
  only, with only combinatoric ambiguities included.} 
\vspace*{-.3in} 
\end{center} 
\end{figure} 
 
1. {\bf Wrong combinations.} For a given event a 
``combination'' is a particular assignment  
of the jets and leptons to the external legs of Fig.~\ref{fig:topology}. 
For each event, there is 
only one correct combination (excluding $1357 
\leftrightarrow 2468$ symmetry).  Assuming that we can 
identify the two jets that correspond to the two quarks, we have 8 
(16) possible combinations for the $2\mu2e$ ($4\mu$ or $4e$) channel. 
The total number of combinations for a pair of events is the product 
of the two, \ie\ 64, 128 or 256. Adding the wrong combination pairings for the 
ideal case yields the mass distributions of Fig.~\ref{fig:allcombis}. 
Compared to Fig.~\ref{fig:ideal}, there are 16 times more (wrong) 
solutions, but the $\delta$-function-like mass peaks remain evident. 
 
2. {\bf Finite widths.} For SPS1a, the widths of the intermediate 
particles are roughly 5~GeV, 20~MeV and 200~MeV for $\sq_L$, $\cntwo$ 
and $\slep_R$.  Thus, the widths are quite small in comparison to the 
corresponding masses.  
 
3. {\bf Mass splitting between flavors.} The masses for up and down type 
squarks have a small difference of 6 GeV. Since it is impossible to 
determine flavors for the light jets, the mass determined should be 
viewed as the average value of the two squarks (weighted by the parton 
distribution functions). 
 
4. {\bf Initial/final state radiation.} These two types of radiation 
not only smear the visible particles' momenta, but also provide a 
source for extra jets in the events.  We will apply a $p_T$ cut to get 
rid of soft jets.    
 
5. {\bf Extra hard particles in the signal events.} In SPS1a, many of 
the squarks come from gluino decay ($\gl\rightarrow q\sq_L$), which yields 
another hard $q$ in the event. Fortunately, for SPS1a  
$m_{\gl} - m_{\sq_L}=40\gev$  is much 
smaller than $m_{\sq_L}-m_{\cntwo}=380\gev$. 
Therefore, the $q$ from squark decay is usually much more energetic 
than the $q$ from $\gl$ decay.  We select the two jets with highest 
$p_T$ in each event after cuts.  Experimentally one would want to 
justify this choice by examining the jet multiplicity to ensure that 
this analysis is dominated by 2-jet events, and not 3 or 4 jet events. 
 
6. {\bf Background events.}  The SM backgrounds are negligible 
for this signal in SPS1a. There are a few significant backgrounds from
other SUSY processes:
 
(a) $\sq_L\rightarrow q\cntwo\rightarrow q\tau\stau\rightarrow 
  q\tau\tau\cnone$ for one or both decay chains, with all $\tau$'s 
  decaying leptonically. Indeed,  $\cntwo\rightarrow\tau\stau$ has the 
  largest partial width, being 14 times that of 
  $\cntwo\rightarrow\mu\smu$.  However, to be included in our 
  selection the two $\tau$'s 
  in one decay chain must both decay to leptons with the same flavor, which 
  reduces the ratio. A cut on lepton $p_T$ also 
  helps to reduce this background, since leptons from $\tau$ decays are 
  softer.  Experimentally one should perform a separate search for 
  hadronically decaying tau's or non-identical-flavor lepton decay chains to 
  explicitly measure this background. 
   
  (b) Processes containing a pair of sbottoms, which have different masses  
from the first two generations.  
 Since $b$ jets are distinguishable, a separate analysis should be 
 performed to determine the $b$ squark masses. However, this presents a 
  background to the light squark search since $b$-tagging efficiency is 
  only about 50\% at high $p_T$. 
 
(c) Processes that contain a pair of $\cntwo$'s, not both 
  coming from squark decays. For these events to fake signal events, 
  extra jets need to come from  
  initial and/or final state radiation or other particle decays. For 
  example, direct $\cntwo$ pair production or $\cntwo+\gl$ 
  production. These are electroweak processes, but, since $\cntwo$ has a 
  much smaller mass than squarks, the cross-section is not negligible. 
  In our SPS1a analysis, the large jet $p_T$ cut reduces this kind of 
  background due to the small $m_{\gl} - m_{\sq_L}$.

7. {\bf Experimental resolutions.} In order to estimate this
experimental effect at the LHC, events in both signal and the
aforementioned SUSY backgrounds are further processed with PGS
\cite{pgs}. Note that in \cite{Cheng:2008mg}, we used ATLFAST for the
detector simulation. Compared with ATLFAST, PGS has more stringent lepton 
isolation cuts, therefore we obtain fewer events. Nevertheless, as shown 
below, the results turn out to be similar.  All objects including 
jets, isolated leptons and missing $p_T$ are taken directly from 
PGS.

The cuts used to isolate the signal are:

I) 4 isolated leptons with $p_T>10$ GeV, $|\eta|<2.5$ and 
matching flavors and charges consistent with our assumed $\cntwo \to 
\slep \to \cnone$ decay; 
 
II) No $b$-jets and $\geq 2$ jets with $p_T>100$ GeV, 
$|\eta|<2.5$. The 2 highest-$p_T$ jets are taken to be particles 7 
and 8;  
 
III) Missing $p_T>50$ GeV. 
 
\noindent For a data sample with 300 $fb^{-1}$ integrated luminosity, there are 
about 620 events left after the above cuts, out of which about  
420 are signal events.  After taking all possible pairs for all possible 
combinations and solving for 
the masses, we obtain the mass distributions in 
Fig.~\ref{fig:smeared}. 

\begin{figure} 
\begin{center} 
 \includegraphics[width=0.6\textwidth]{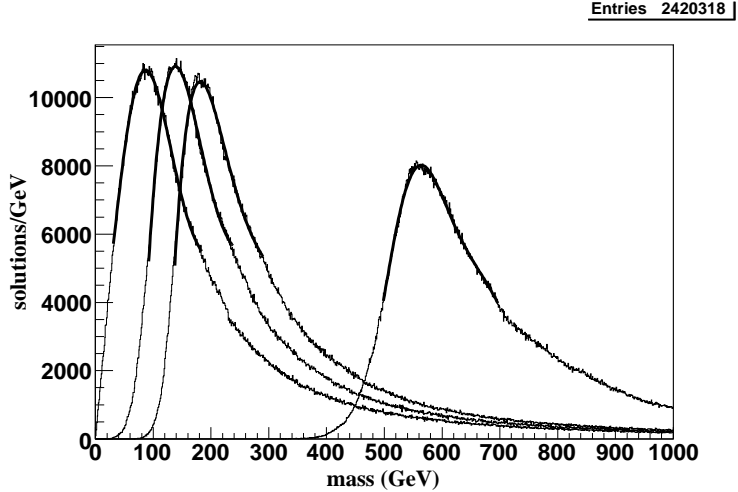} 
\vspace*{-.1in} 
\caption{\label{fig:smeared}Mass solutions with all effects 1 -- 7 
  included and after cuts I 
  -- III for the SPS1a SUSY model and $L=300\fbi$. All effects 
  incorporated, including backgrounds.} 
\vspace*{-.3in} 
\end{center} 
\end{figure} 
{From} Fig.~\ref{fig:smeared}, we see that the mass peaks are smeared 
but still present around the input masses. The analytical formula for 
the distributions are unknown, so we  
estimate the masses by reading the peak positions. To minimize the 
effect from statistical fluctuations, we fit each distribution using a 
sum of a Gaussian plus a (single)  
quadratic polynomial and taking the maximum positions of the fitted 
peaks as the estimated masses. We will use this function as the 
``standard fit'' throughout this article. The fitted range is
restricted to be above the half height. The fitted curves are
superimposed on the mass distributions in Fig.~\ref{fig:smeared},
which yields \{78.4, 134.2, 181.5, 553.9\}$\gev$ for  
the masses.  Averaging over 20 different data samples, we find  
\bea  
&&m_N=76.7\pm2.0\gev,\quad 
m_X=134.6\pm2.2\gev,\cr  
&&m_Y=178.9\pm3.8\gev,\quad m_Z=561.6\pm5.4\gev.\label{eq:sps1a_before} 
\eea  
The statistical uncertainties are very small, but 
there exist biases, especially for the two light masses. In practice, 
we can always correct the biases by comparing real data with Monte 
Carlo. Nevertheless, we would like to reduce the biases as much as 
possible using data only. In some cases, the biases can be 
very large and it is essential 
to reduce them before comparing with Monte Carlo--we will see an 
example later. 
 
The combinatorial background is an especially important source of bias 
since it yields peaked mass distributions that are 
not symmetrically distributed around the true masses, 
as can be seen from Fig.~\ref{fig:allcombis}. This will introduce 
biases that survive even after smearing. Therefore, we concentrate on 
reducing wrong solutions.   
 
\begin{figure} 
\begin{center} 
 \includegraphics[width=0.8\textwidth]{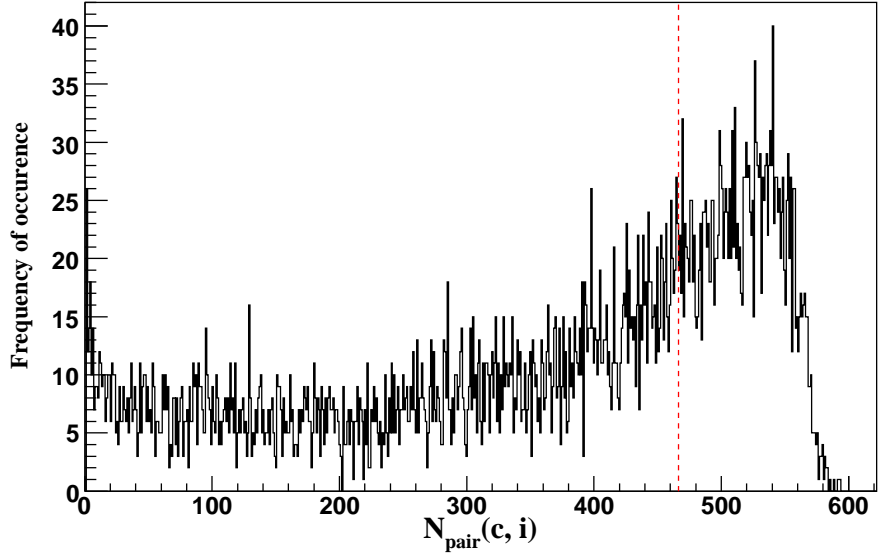} 
\caption{\label{fig:nevents}For each event, $i$, and each combination, 
  $c$, associated with that event, we count the number, 
  $N_{pair}(c,i)$, of events that can pair with it and give at least 
  one solution. The plot shows the frequency of occurrence of different 
  values of $N_{pair}(c,i)$. All effects are incorporated, including 
  backgrounds.  The plot is for the SPS1a case (for which the total 
  number of signal+background events is 620 for $L=300\fbi$).  In the 
  bias reduction procedure, any choices of $c,i$ yielding 
  $N_{pair}(c,i)$ to the left of the red line (corresponding to 75\% 
  of the total number of events) are discarded.} 
\end{center} 
\end{figure}

First, we reduce the number of wrong combinations by the following 
procedure. For each combination choice, $c$, for a given event, $i$ 
($i=1,N_{evt}$), we count the number, $N_{pair}(c,i)$, of events that 
can pair with it (for some combination choice for the 2nd events) and 
give us solutions. We repeat this for every combination choice for 
every event. Neglecting effects 2.-- 7., $N_{pair}(c,i)=N_{evt}-1$ if 
$c$ is the correct combination for event $i$.  After including 
backgrounds and smearing, $N_{pair}(c,i)<N_{evt}-1$, but the correct 
combinations still have statistically larger $N_{pair}(c,i)$ than the 
wrong combinations. The frequency with which various values of 
$N_{pair}(c,i)$ occur is shown as a function of $N_{pair}(c,i)$ in 
Fig.~\ref{fig:nevents}.

 To enhance the likelihood that a particular choice of $c,i$ 
 corresponds to a correct solution, we cut on $N_{pair}(c,i)$. 
For the SPS1a model point, if $N_{pair}(c,i)\leq 0.75\,N_{evt}$ we 
discard the combination choice, $c$, for event $i$.  If all possible 
$c$ choices for event $i$ fail this criterion, then we discard event 
$i$ altogether (implying a smaller $N_{evt}$ for the next analysis 
cycle). We then repeat the above procedure for the remaining events 
until no combinations can be removed. After this, for the example data 
sample, the number of events is reduced from 622 (424 signal + 198 
background) to 430 (322 signal + 108 background), and the average 
number of combinations per event changes from 11 to 4. 
 
Second, we increase the significance of the true solution by weighting 
each surviving pair of events by $1/n$ where $n$ 
is the number of solutions for the given pair (using only the 
combination choices that have survived the previous cuts).  This causes 
each pair (and therefore each event) to have equal weight in our 
histograms.  Without this weighting, a pair with multiple solutions
has more weight than a pair with a single solution, even though at
most one solution would be correct for each pair.

Finally, we exploit the fact that wrong solutions and 
backgrounds are much less likely to yield $M_N, M_X, M_Y$, and $M_Z$ 
values that are all simultaneously close to their true values. We plot 
the $1/n$-weighted number of solutions as a function of the three mass 
differences (Fig.~\ref{fig:dmass}).  We define mass difference windows 
by $0.6\times $(peak height) and keep only those solutions 
for which {\it all three} mass differences fall within the mass 
difference windows. 
\begin{figure} 
\begin{center} 
 \includegraphics[width=0.6\textwidth]{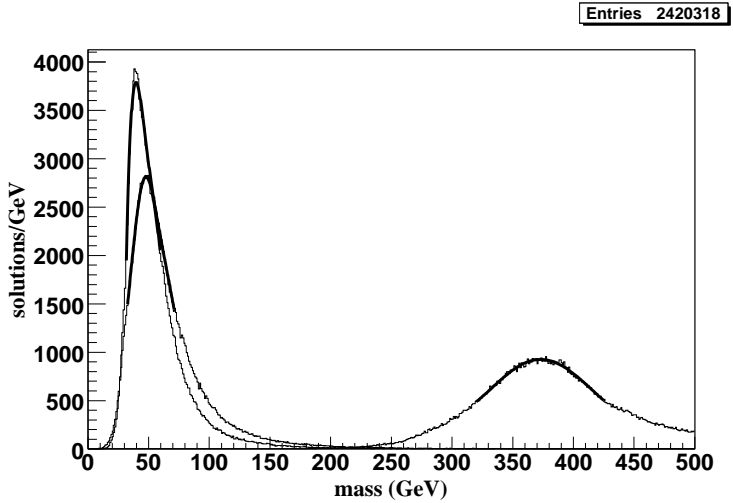} 
\vspace*{-.1in} 
\caption{\label{fig:dmass}SPS1a, $L=300\fbi$ mass difference 
  distributions. All effects incorporated, including backgrounds. } 
\vspace*{-.3in} 
\end{center} 
\end{figure} 
\begin{figure}[t!] 
\begin{center} 
 \includegraphics[width=0.6\textwidth]{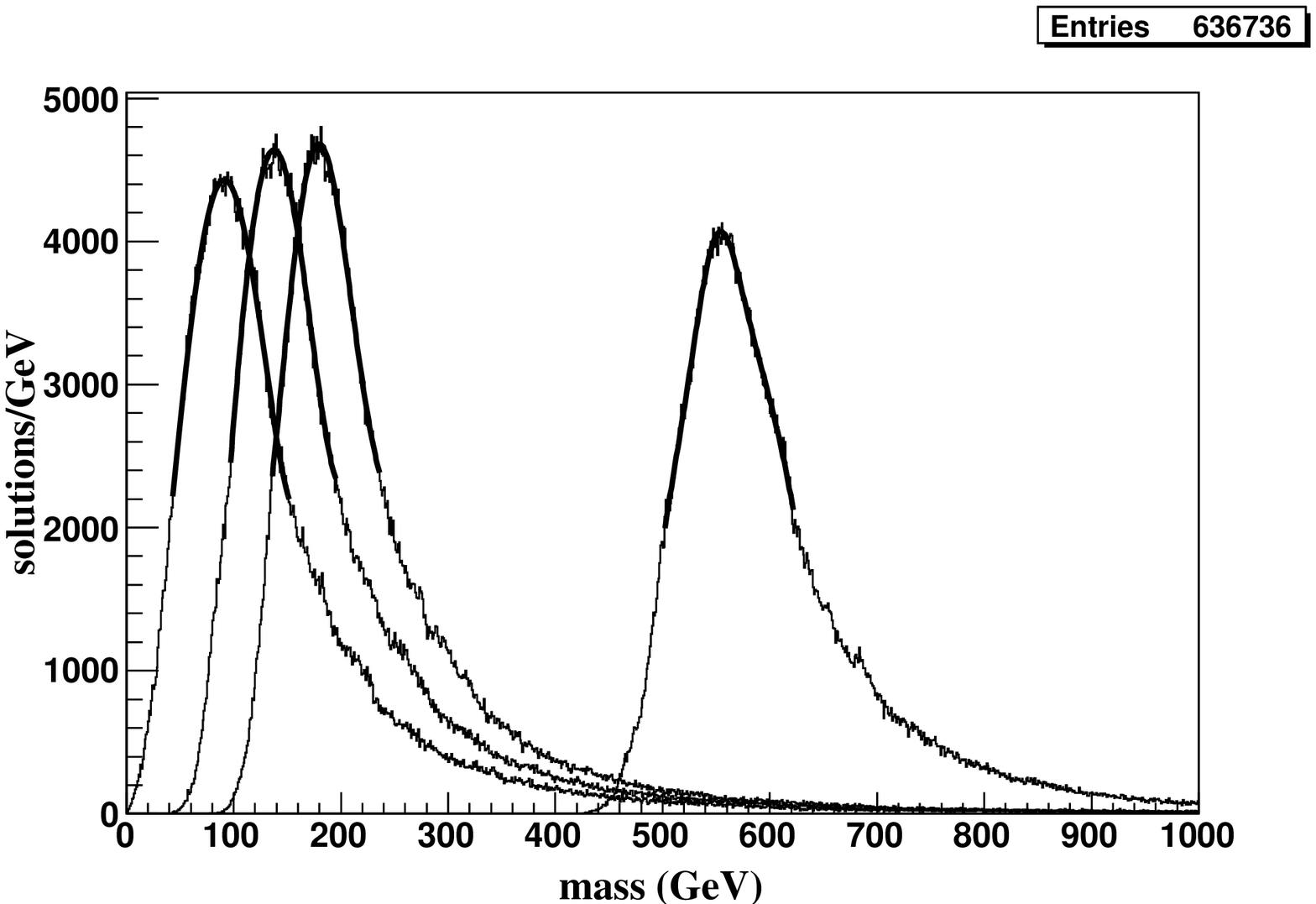} 
\vspace*{-.1in} 
\caption{\label{fig:masses_final}Final mass distributions after 
  the bias reduction procedure for the SPS1a SUSY model and 
  $L=300\fbi$. All effects incorporated, including backgrounds.} 
\vspace*{-.3in} 
\end{center} 
\end{figure} 
The surviving solutions are plotted (without the $1/n$ weighting) in 
Fig.~\ref{fig:masses_final}. Compared 
with Fig.~\ref{fig:smeared}, the mass peaks are narrower, more 
symmetric and the fitted values are less biased. The fitted masses 
are \{93.9, 140.3, 180.5, 559.2\} GeV. Repeating the procedure for 20 
data sets, we find 
\bea 
&&m_N=93.8\pm3.9\gev, \quad m_X=138.4\pm4.5\gev,\cr 
&&m_Y=178.7\pm4.6\gev,\quad m_Z=559.5\pm5.4\gev \,,\label{eq:sps1a_results} 
\eea 
to be compared to the input masses of Eq.~(\ref{minsps1a}). 
Thus, the biases are reduced without significantly increasing the 
statistical errors.

Thus, we have shown that the masses can be measured with high precision 
for a few hundred events in the 4-fermion decay channel. In the case 
of the SPS1a point, the number of events employed above corresponds to 
a high integrated luminosity, $L\sim$ 300 $fb^{-1}$. The reason that 
such a high luminosity is required in the case of the SPS1a scenario 
is that the branching ratio for $\cntwo\rightarrow\stau\tau$ is 14 
times that for $\cntwo\rightarrow\smu\mu$ or $\cntwo\rightarrow\se 
e$. More generally, the integrated luminosity needed to get 
a few hundred events is highly dependent on the branching ratios for 
the various SUSY particle decays in the model. For example, if one 
takes the SPS1a masses but requires that $\cntwo$ decays equally to 
the three lepton flavors instead, the same number of signal events as 
employed above can be obtained with just 10 $fb^{-1}$ of data.  
 
Although the errors in the mass determinations depend upon the number 
of events, our method is quite robust in that we get decent mass 
determinations even with a small number of events. In 
Fig.~\ref{fig:50events}, the mass distributions for 50 events are 
shown, with evident mass peaks. By repeating our  procedure for 
multiple  datasets of a given size, we obtain the errors as functions
of the number of events. Fig.~\ref{fig:error} shows the error for the
$\cnone$ mass determination as a function of the number of
signal+background events. Note that the central value for multiple
data sets of the given size is quite insensitive to the data set size,
but, of course, the possible deviation from this central value for any
one data set increases as the data set size decreases. 
\begin{figure} 
\begin{center} 
 \includegraphics[width=0.6\textwidth]{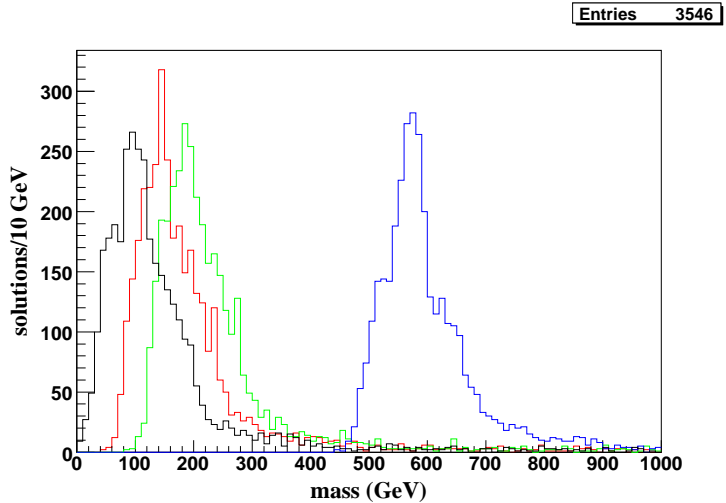} 
\vspace*{-.1in} 
\caption{\label{fig:50events}Mass distributions for 50 events for SPS1a.} 
\vspace*{-.3in} 
\end{center} 
\end{figure} 
\begin{figure} 
\begin{center} 
 \includegraphics[width=0.6\textwidth]{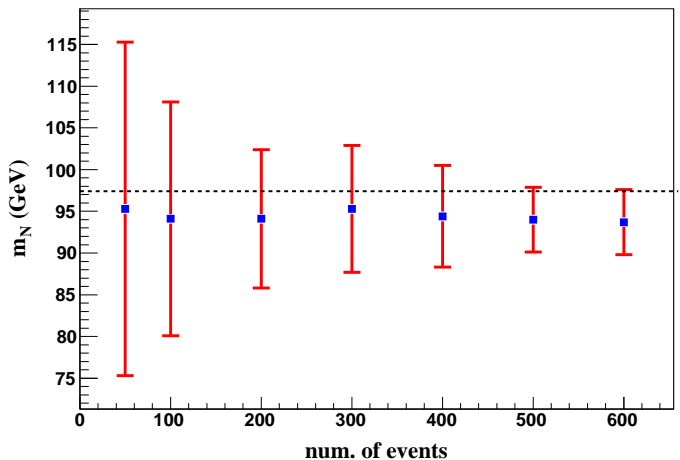} 
\vspace*{-.1in} 
\caption{\label{fig:error}Error bars for $m_N$  
  as a function of the number of background+signal events, for
  SPS1a. All effects and procedures included.} 
\vspace*{-.3in} 
\end{center} 
\end{figure}

\subsection{SUSY Point \#1} 
 
We have applied our method to other mass points to show its 
reliability. We quote here results for ``point \#1'' defined in 
Ref.~\cite{Cheng:2007xv} with the following masses: \{85.3, 128.4, 
246.6, 431.1/438.6\} GeV. For 100 fb${}^{-1}$ data, we have about 800 
events (770 signal events) after the same pre-bias-reduction cuts. 
The resulting mass plot before performing bias reduction cuts is that 
given in Fig.~\ref{point1before}. From Fig.~\ref{point1before}, we see 
that the mass peaks are very broad and we get more than 50 GeV biases 
if we use the positions of the maxima as the true mass values. We then 
repeat the same bias reduction procedure as for SPS1a except that we 
employ a looser cut on $N_{pair}(c,i)$ than for the SPS1a case,
despite the fact that there are more signal events for Point \#1. We
require $N_{pair}(c,i)>0.6\,N_{evt}$. The reason is that,  
unlike the SPS1a case, the gluino mass in Point \#1 (524 GeV) is 
significantly larger than the squark mass. Therefore the quark jet 
from gluino decay is often misidentified as the jet from squark decay, 
which reduces the chance to obtain solutions for a pair of 
events.~\footnote{It is possible to improve the results by considering 
  all high $p_T$ jets as candidates for the quarks from squark decays, 
  instead of simply choosing the two highest $p_T$ jets.}  In 
practice, there is not a universal ``best'' cut on $N_{pair}(c,i)$: a 
more stringent cut leads to smaller biases but larger statistical 
uncertainties. After the bias reduction procedure 
using $N_{pair}(c,i)>0.6\,N_{evt}$ we are left with 560 
events (550 signal events). The mass distributions are shown in 
Fig.~\ref{point1after}. They are much narrower and the biases are 
considerably reduced. After following the bias reduction procedure and 
using 20 data samples to estimate the errors, we obtain 
$m_N=82.8\pm3.2\gev$, $m_X=127.9\pm3.0\gev$, $m_Y=245.7\pm3.4\gev$, 
$m_Z=436.4\pm 5.4\gev$. The central values are in quite close 
agreement with the input masses 
except for $m_N$ which comes out a bit low.   
 
\begin{figure} 
\begin{center} 
 \includegraphics[width=0.6\textwidth]{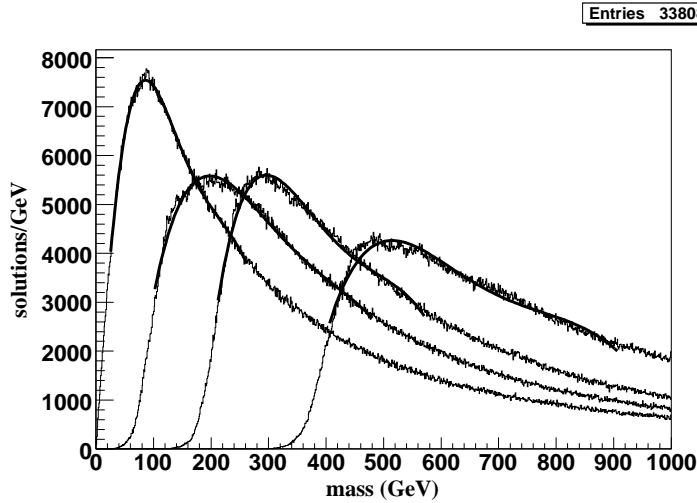} 
\vspace*{-.1in} 
\caption{\label{point1before}Final mass distributions before 
  the bias reduction procedure for the point \#1 SUSY model and 
  $L=100\fbi$. All effects incorporated, including backgrounds.} 
\vspace*{-.3in} 
\end{center} 
\end{figure} 
\begin{figure} 
\begin{center} 
 \includegraphics[width=0.6\textwidth]{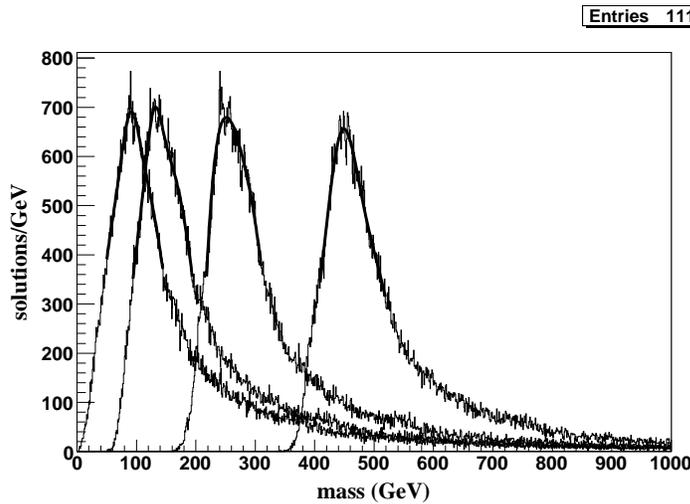} 
\vspace*{-.1in} 
\caption{\label{point1after}Final mass distributions after 
  the bias reduction procedure for the point \#1 SUSY model and 
  $L=100\fbi$. All effects incorporated, including backgrounds.} 
\vspace*{-.3in} 
\end{center} 
\end{figure} 
 
\subsection{Comments and Comparisons} 
 
We emphasize that the remaining biases in the above mass determinations can 
be removed by finding those input masses that yield the observed 
output masses after processing Monte Carlo generated data through our 
procedures. In this way, very accurate central mass 
values are obtained with the indicated statistical errors. 
 
The above results for the $N$, $Y$ and $X$ masses for  
the SPS1a point and point \#1 can be compared to those obtained  
following the very different procedure of 
Ref.~\cite{Cheng:2007xv}. There, only the $X\to Y \to N$ parts of the 
two decay chains were employed and we used only $4\mu$ events. For the 
SPS1a model point we obtained $m_N=98\pm9\gev$, $m_Y=187\pm10\gev$, and 
$m_X=151\pm 10\gev$. And, for point \#1 we found $m_N=86.2\pm 
4.3\gev$, $m_X=130.4\pm 4.3\gev$ and $m_Y=252.2\pm 4.3\gev$. Including 
the $4e$ and $2\mu 2e$ channels will reduce the indicated errors by a factor 
of $\sim 2$. The procedure of \cite{Cheng:2007xv} can thus be used to 
verify the results for $m_N$, $m_X$ and $m_Y$ from the present 
procedure and possibly the two can be combined to obtain smaller 
errors than from either one, with $m_Z$ determined by the procedure of 
this paper. 
 
We also compare the results for SPS1a with those given in 
Ref.~\cite{Cheng:2008mg} where exactly the same procedure and cuts are 
applied to the same model point. The difference is that we used 
ATLFAST for the detector simulation in Ref.~\cite{Cheng:2008mg} while 
we have switched to PGS in the current paper. The PGS simulation has more 
stringent lepton isolation cuts and therefore we obtain fewer events 
in the present analysis (620 {\it vs} 1050). In 
Ref.~\cite{Cheng:2008mg}, we obtained 
\bea  
&&m_N=76.7\pm1.4\gev,\quad 
m_X=135.4\pm1.5\gev,\cr  
&&m_Y=182.2\pm1.8\gev,\quad m_Z=564.4\pm2.5\gev.\nonumber 
\eea  
before the bias reduction procedure and  
\bea 
&&m_N=94.1\pm2.8\gev, \quad m_X=138.8\pm2.8\gev,\cr 
&&m_Y=179.0\pm3.0\gev,\quad m_Z=561.5\pm4.1\gev.\nonumber 
\eea   
after. Comparing the above numbers with those in 
Eqs.~(\ref{eq:sps1a_before}) and (\ref{eq:sps1a_results}), we see that 
the masses obtained using PGS simulation have larger statistical 
errors, in accord with the smaller number of events. On the other 
hand, the central values agree well, indicating that the bias 
reduction procedure affects the mass peaks in a nearly 
model-independent manner.  We view this as evidence of the robustness 
of our method. 
 
\subsection{Removing Biases Using a Dilepton Edge Cut} 

As we have discussed, the primary source of biases is the detector 
smearing of wrong solutions, especially those associated with wrong 
combinations. It will be possible to efficiently eliminate many of 
these wrong solutions if there is a significant structure associated 
with correct solutions in one or more distributions constructed from 
the visible particles' momenta. In the SUSY examples we consider here, 
such a structure is especially apparent in the distribution of $\mll$, 
where $\ell=e,\mu$ (same flavor pairs only). The advantage of using 
only leptons is the much better resolution for the lepton momentum 
measurements.  Ignoring resolution smearing, kinematics predicts that 
correct combinations should have 
\begin{equation} 
(\mll^{\rm edge})^2=\frac{(m_{Y0}^2-m_{X0}^2)(m_{X0}^2-m_{N0}^2)}{m_{X0}^2},\label{eq:edge} 
\end{equation} 
where $m_{Y0}$, $m_{X0}$ and $m_{N0}$ are the input masses.  Note that 
there are many more dilepton events than four lepton events since 
dileptons require only a single decay chain of 
Fig.~\ref{fig:chain_decay}. 
The plot of $\mll$ values for all solutions coming from 600 SPS1a 
events (after PGS smearing and general cuts, but before applying the 
bias reduction procedure) is shown in Fig.~\ref{fig:mllsps1a}.  The edge at 
the predicted value of $80\gev$ is apparent and its location can be 
determined quite accurately from the data. 
 
\begin{figure} 
\begin{center} 
 \includegraphics[width=0.6\textwidth]{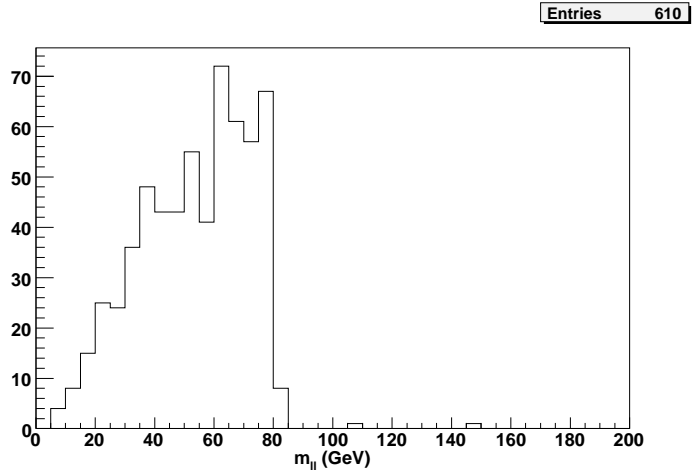} 
\vspace*{-.1in} 
\caption{\label{fig:mllsps1a}We plot the number of events as a 
  function of $\mll$ for 600 
  SPS1a events in 5 GeV bins (after PGS smearing 
  and general cuts, but before the bias reduction procedure). Only events 
  containing two muons and two electrons with opposite charges are used to avoid 
  ambiguity, each of which conntributes two entries to the 
  histogram. The edge at 80 GeV is apparent.  
 } 
\vspace*{-.3in} 
\end{center} 
\end{figure} 
 
Before employing the bias reduction procedure, we apply a cut on the 
$m_Y,m_X,m_N$ values obtained for a given solution of 
\begin{equation} 
\left|\sqrt{(m_Y^2-m_X^2)(m_X^2-m_N^2)/{m_X^2}}-\mll^{\rm 
  edge}\right|<20\gev, \label{eq:mllcut} 
\end{equation} 
where we have purposely employed a rather loose cut so as to not lose 
statistics and to take into account smearing of the input $X,Y,N$ 
masses that will be present even for a correct combination, as well as 
the small error associated with determining the edge location experimentally. 
We {\it then} use the same bias reduction procedure as discussed 
earlier using a sequence of choices for the cut $f_{cut}$ defined by 
retaining only combinations with $N_{pair}(c,i)>f_{cut}N_{evt}$. In Table~\ref{diltable}. 
 
\begin{table}[htb] 
\begin{center} 
{\footnotesize 
\begin{tabular}{c||c|c|c|c||c|c|c|c} 
\hline 
&\multicolumn{4}{c||}{with dilepton edge cut}&\multicolumn{4}{c}{without dilepton edge cut} 
\\ 
\hline 
$f_{cut}$ & 0.60 & 0.65 & 0.70 &0.75 & 0.60 & 
0.65 & 0.70 & 0.75\\ 
\hline 
$m_N$ (GeV) & 
$93.0\pm3.7$&$96.1\pm3.9$&$97.5\pm4.3$&$97.9\pm4.9$&$85.6\pm2.3$&$88.1\pm3.5$&$90.7\pm3.8$&$93.8\pm3.9$\\ 
\hline 
$m_X$ (GeV)&$138.9\pm3.9$&$141.4\pm4.6$&$143.7\pm4.6$&$144.3\pm4.0$&$131.5\pm2.7$&$133.9\pm3.6$&$135.9\pm4.3$&$138.4\pm4.5$\\ 
\hline 
$m_Y$ (GeV)&$176.5\pm3.8$&$178.8\pm4.6$&$180.8\pm5.1$&$181.5\pm5.3$&$172.8\pm2.8$&$174.8\pm3.8$&$176.6\pm4.4$&$178.7\pm4.6$\\ 
\hline 
$m_Z$ (GeV)&$557.8\pm4.4$&$559.9\pm4.5$&$563.2\pm5.0$&$565.6\pm6.2$&$555.8\pm5.2$&$557.2\pm5.5$&$557.8\pm5.1$&$559.5\pm5.4$\\ 
\hline 
\end{tabular} 
} 
\end{center} 
\caption{Peak locations for various values of $f_{cut}$ with and 
  without the dilepton edge cut. Errors were determined using 20 
  distinct data sets.\label{diltable}} 
\end{table} 
We clearly observe that the dilepton edge cut has greatly reduced the 
bias in comparison to results obtained without the dilepton edge 
cut. Further, for the larger values of $f_{cut}$, the mass peak 
locations are not biased at all (within statistics) in comparison to 
the input masses of Eq.~(\ref{minsps1a}). This occurs because many of 
the wrong solutions have been elliminated. For example, after the 
dilepton edge cut and after employing the bias-reduction procedure using  
$f_{cut}=0.75$,  about 160 events are retained on average and the average 
number of solutions for the remaining pairs formed from these 
surviving events is only about 1.2. In other words, the dilepton edge 
cut is highly effective in removing wrong combinations. Errors 
for the peak mass values are, of course, slightly larger when a 
dilepton edge cut is imposed, implying that ultimately the best mass 
determinations may be those obtained using Monte Carlo determination 
of the bias corrections that should be applied to mass peak values 
obtained without the dilepton edge cut.  Nonetheless, doing the 
analysis with a dilepton edge cut will provide a very important cross 
check of the bias determination. 
 
As a final note, we observe that in the case of SPS1a there are also 
incorrect solutions coming from chains containing a pair of 
leptonically decaying $\tau$'s.  Many, but not all of these wrong 
solutions are also be eliminated by the dilepton edge cut. The 
remaining background events contain mostly those events for which one 
chain has $\ell=e,\mu$ while the other has a pair of leptonically 
decaying $\tau$'s, since sometimes such events will give a solution 
with nearly correct mass values.

\subsection{More on Backgrounds} 
 
Because the SM background can be efficiently reduced by applying a 
large missing $p_T$ cut, the most difficult backgrounds usually come from 
other SUSY processes that contain the same final state particles. In 
the above examples, we have already encountered such backgrounds. In 
the SPS1a case, the backgrounds are dominated by events that contain 
leptonically decaying $\tau$'s. For SUSY point \#1, although 
most events are originally signal events, in many cases the jets from 
squark decays are not correctly identified, in which case these events 
should be viewed as background events. In both examples, the 
background events are closely related to the signal events and 
therefore also carry some information about the masses. It is also 
interesting to study the effects of background events of a completely 
different origin, and test the stability of our mass determination method.  
 
\begin{figure} 
\begin{center} 
 \includegraphics[width=0.6\textwidth]{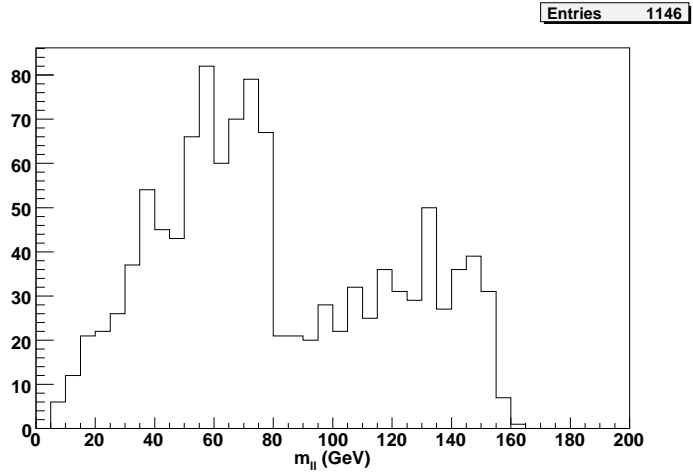} 
\vspace*{-.1in} 
\caption{\label{fig:mll}Dilepton invariant mass distribution for 600 
  SPS1a events together with 600 Point 1 events (after PGS smearing 
  and general cuts, but before the bias reduction procedure). Only events 
  containing two muons and two electrons of opposite charge are used to avoid 
  ambiguity. The two edges at 80 GeV and 157 GeV correspond to SPS1a 
  and Point \#1 respectively. } 
\vspace*{-.3in} 
\end{center} 
\end{figure} 
 
In order to explore the issues that arise, we will perform an analysis 
in which we consider the SPS1a events as signal events, fixing the 
number of events to 600 (including the intrinsic background of SPS1a). 
For a possible SUSY background to the SPS1a events we employ events of 
the above SUSY Point \#1 as ``background'' events, varying the ratio 
of Point \#1 events with respect to the SPS1a events. Since this is 
only for illustration, we are not concerned with how this could happen 
in a specific SUSY model. The existence of two different type of 
events is immediately seen from the dilepton invariant mass 
distribution (Fig.~\ref{fig:mll}), where two different edges are 
evident. The position of the edges are given by Eq.~(\ref{eq:edge}). 
In this case, the two signals give two different $m_{Y0}$, $m_{X0}$ 
and $m_{N0}$ input mass sets. Again, there are many more dilepton 
events than four lepton events since a dilepton only requires a single 
decay chain of Fig.~\ref{fig:chain_decay}. From Fig.~\ref{fig:mll} we 
see that the position of the edge associated with the SPS1a signal can 
be determined quite precisely even when the background to signal ratio 
is of order one.  Consequently, one can try to combine the edge 
location measurement with information from double chain events (see 
also Ref.~\cite{hybrid}). 
 
First, we repeat our fitting procedure on the mixed events without 
using the dilepton edge information. Since there are more background 
events, we cannot use the fixed $N_{pair}(c,i)>0.75\,N_{evt}$ cut as 
before. This is because $N_{evt}$ now refers to the total number of 
events from both SPS1a and SUSY Point \#1 so that such a cut would 
amount to a much stronger $f_{cut}$ value for the SPS1a signal on its 
own.  Instead, we choose the $N_{pair}(c,i)$ cut so that 60\% of all 
the events are left after the bias reduction procedure. The 
corresponding value of $f_{cut}$ varies according to the amount of 
SUSY Point \#1 background included.  For 600 SPS1a events combined 
with 600 Point \#1 background events the 60\% survival fraction 
corresponds to using $f_{cut}\sim 0.58$ in the bias-reduction procedure. 
For the SPS1a signal alone, the corresponding $f_{cut}$ 
value is somewhat larger. The measured $m_N$ is shown in 
Fig.~\ref{fig:mixed_error} as a function of the number of background 
events (after PGS smearing, but before the bias reduction procedure). 
 \begin{figure} 
\begin{center} 
 \includegraphics[width=0.6\textwidth]{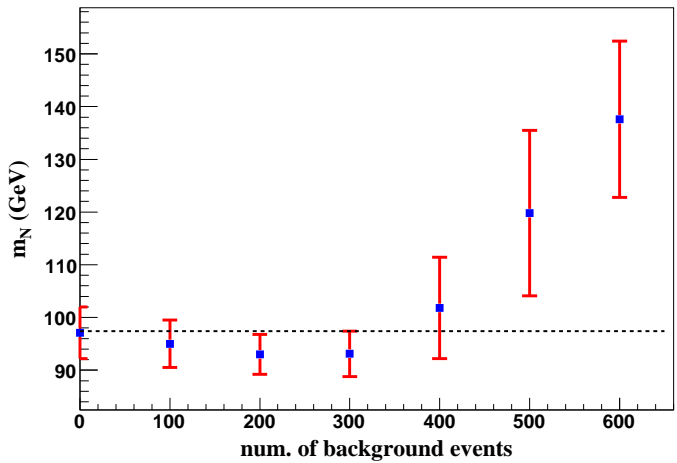} 
\vspace*{-.1in} 
\caption{\label{fig:mixed_error} The measured $m_N$ as a function of 
  the number of background events. The number of signal (SPS1a) events 
is fixed to 600.  In both cases, these are the event numbers after 
general cuts but before the bias reduction procedure.} 
\vspace*{-.3in} 
\end{center} 
\end{figure} 
{From} Fig.~\ref{fig:mixed_error}, we see that the mass determination is 
not accurate when the background/signal ratio is high. However, as 
long as the number of background events after general cuts is less 
than about half the number of signal events, the bias reduction 
procedure is effective in removing background events while retaining 
signal events, and the mass determination is quite good. Of course, in 
practice we will not know a priori what the number of background 
events is relative to the number of signal events and therefore we 
would need additional input in order to know if the fitted masses from 
mass peaks are reliable estimates for the true masses. In the present 
case, the dilepton mass plot of Fig.~\ref{fig:mll} would clearly have 
indicated the presence of two different classes of events and we would 
therefore know that it would help to use an additional cut to reduce 
the ``background'' class. We again employ the simple cut of 
Eq.~(\ref{eq:mllcut}), where in order to isolate the SPS1a component 
of the combined events we would employ $\mll^{\rm edge} = 80\gev$. 
Again note that this is a loose cut that does not require a precise 
knowledge of the edge position from the single chain events. Applying 
the cut in Eq.~(\ref{eq:mllcut}) on datasets with 600 signal + 600 
background events (after general cuts, but before the bias reduction 
procedure) and repeating the fit procedure, we 
obtain the masses 
\bea 
m_N&=&96.2\pm 3.6\gev, \quad m_X=141.3\pm4.1\gev, \cr m_Y&=&178.4\pm 
4.1\gev,\quad m_Z=558.5\pm 4.6\gev\,, 
\label{lledgemasses} 
\eea 
where errors were determined using 20 distinct data sets.  The same 
$f_{cut}=0.58$ was used as in the case without dilepton edge cut. 
With the dilepton edge cut, we are left with averagely 
363 SPS1a events and 8 SUSY Point \#1 events in the mass 
distributions used to get the mass peak locations of 
Eq.~(\ref{lledgemasses}). Thus, we effectively isolated the signal of 
interest by employing the dilepton edge cut. We have also obtained 
central mass values that have almost no bias relative to the input 
SPS1a masses. This is because the $f_{cut}$ for the SPS1a signal alone 
that yields the number of events ($\sim 360$) after bias reduction is 
close to $0.65$, which according to Table~\ref{diltable} should give a 
nearly unbiased mass determination.   
 
\section{Comparison of SUSY and UED} 
\label{sec:UED} 
 
In this section, we address the question of whether or not the mass 
determinations (and the accuracy thereof) are sensitive to the model 
employed by comparing results for the SPS1a point to a UED model 
chosen so that the masses are exactly the same as for SPS1a (the 
corresponding decay chain in UED is: KK-quark $\rightarrow$ KK-$Z$ 
$\rightarrow$ KK-lepton $\rightarrow$ KK-photon). We have also adjusted 
the squarks/KK-quarks of different flavors to have the same mass 
(564.8 GeV, as for $\wt{u}_L$ in SPS1a) and chosen squark/KK-quark 
pair production as the only process ({\it i.e.}, no gluinos/KK-gluons). 
The finite widths of the involved particles are also turned off.  
Both SUSY and UED events are simulated with Herwig++ \cite{Herwig++}, 
with spin correlations included and confirmed by comparing with 
Ref.~\cite{Smillie:2005ar}. In Figs.~\ref{susysps1asignal} and 
\ref{uedsps1asignal}, for SUSY and UED respectively, we plot the mass 
distributions after employing identical smearing (PGS), general cuts 
and the bias reduction procedure (using $N_{pair}(c,i)>0.75\,N_{evt}$). The size of 
the event samples for SUSY and UED are set by requiring that both 
samples contain 400 events after PGS smearing and general cuts, but 
before the bias reduction procedure. 
Visually, it is clear that the peaks are in very similar locations. 
 
\begin{figure} 
\begin{center} 
 \includegraphics[width=0.6\textwidth]{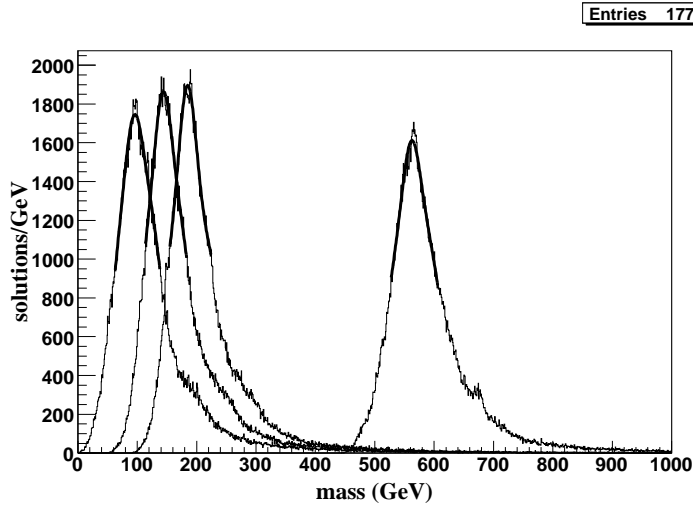} 
\vspace*{-.1in} 
\caption{\label{susysps1asignal}Final mass distributions for signal 
  events only after employing the bias reduction procedure for the SPS1a mass 
  choices in the context of the SUSY model and using 400 signal events 
  after PGS smearing and general cuts but before bias reduction.  
  All combinations and solutions are included, but backgrounds 
  associated with the SUSY model are not included. 
}  \vspace*{-.3in} 
\end{center} 
\end{figure} 
\begin{figure} 
\begin{center} 
 \includegraphics[width=0.6\textwidth]{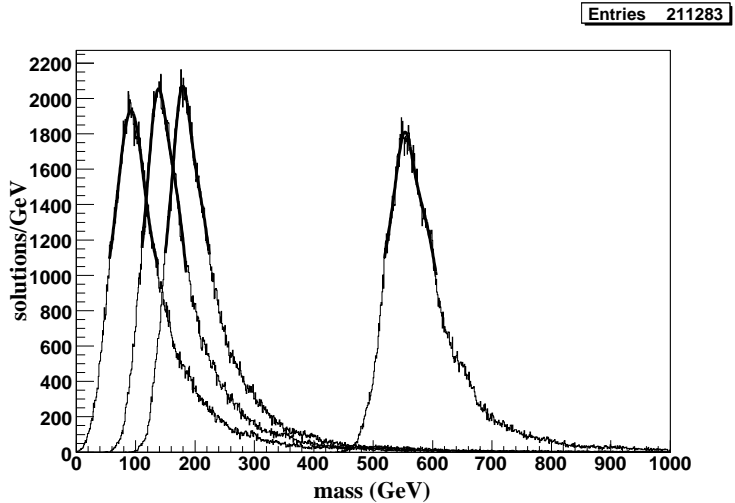} 
\vspace*{-.1in} 
\caption{\label{uedsps1asignal}Final mass distributions for signal events 
  only after the bias reduction procedure for the SPS1a mass choices, 
  but in the case of the UED model. The UED event sample is scaled so 
  that there are 400 signal events after PGS smearing and general cuts 
  but before bias reduction. All combinatorics and solutions are 
  included, but backgrounds associated with the UED model are not included.} 
\vspace*{-.3in} 
\end{center} 
\end{figure} 
 
After employing our standard fitting techniques, we obtain masses of  
\bea  
&&m_N=90.3\pm3.0\gev,\quad 
m_X=135.4\pm3.1\gev,\cr  
&&m_Y=176.0\pm2.9\gev,\quad m_Z=551.2\pm5.2\gev.\nonumber 
\eea  
for the actual SUSY SPS1a model and masses of 
\bea  
&&m_N=89.5\pm3.5\gev,\quad 
m_X=134.8\pm3.5\gev,\cr  
&&m_Y=176.2\pm3.7\gev,\quad m_Z=548.7\pm5.8\gev.\nonumber 
\eea  
for the UED model with SPS1a masses.  We note that the measured masses 
still have some biases of order a few GeV (for the three smaller 
masses) up to more than 10 GeV. But the biases are very similar for the 
two models with different spins, indicating that the major part of these biases 
can be removed by comparing with Monte Carlo even before the 
underlying model is determined. Using the technique of repeating the 
procedure 20 times, we find that the SUSY model statistical errors in 
this case are somewhat smaller than quoted earlier in 
Eqs.~(\ref{eq:sps1a_results}) because we have not included any 
background from other SUSY processes. The biases are also slightly 
different from Eqs.~(\ref{eq:sps1a_results}) where the backgrounds are 
included.  The UED model errors would also be increased by including 
backgrounds coming from other UED processes. In other words, the 
masses are very well determined (to within a few GeV) by our purely 
kinematic procedures, but the errors are mildly model-dependent 
because of variation in the nature and magnitude of the 
new-physics-model backgrounds. The contribution from the backgrounds 
can often be inferred from real data and subtracted. For example, the 
SPS1a background is dominated by the chain decays that yield stau's, whose existence 
and production rate can be determined from hadronically decayed tau's. 
 
\section{Discussion and Conclusions} 
\label{sec:discussion} 
 
One important question is whether it is better to use one-chain or 
two-chain techniques. Our point of view is that one should use all 
available kinematic information, regardless of whether it is from 
one-chain or two-chain events. On the one hand, due to the 
fact that not all events contain two identical chains, one often 
obtains more one-chain events of a certain type than events with two
identical decay chains of that type. However, if one considers only one chain at a time, 
information, in particular that related to the measured missing 
transverse momentum, is always lost.  The consequence is that either 
one cannot solve directly for all involved masses for a given length of decay 
chain, or one must employ longer decay chains, in which case the 
method becomes very complicated.  For example, in the one-chain case 
one needs to employ decay chains with four visible particles (vs. 
three visible particles in the two-chain case) and, in addition, one 
needs to combine five events to obtain discrete solutions for the 
unknown masses.  Even assuming that such events do exist, there are
more wrong combinations and wrong solutions than the two-chain case
studied in this paper. Further, the existence of a certain
type of decay chain implies that there are always events with two 
identical such decay chains. Events with two identical decay chains
always provide more information for the masses of the particles in
the decay chains. The challenge of two-chain techniques is that 
one needs to identify those events in which there are indeed two 
identical chain decays.  Ideally, one would divide the observed events 
into different channels according to their event topologies (chain 
type 1 + chain type 1, chain type 1 + chain type 2, chain type 2 + 
chain type 2, \ldots), apply methods appropriate to each topology and, 
in order not to lose statistics, combine all these channels in the 
analysis. For this reason, it is very important to extend the studies 
in Ref.~\cite{Cheng:2007xv} and the current paper to other event 
topologies. 
 
The importance of using one-chain decay information is illustrated in 
the SPS1a case.  Since there are many more dilepton events than 
4-lepton events the dilepton edge given in Eq.~(\ref{eq:edge}) can be 
measured very precisely. If available, one should certainly 
incorporate this measurement into the two-chain techniques to better 
determine the masses. This kind of ``hybrid'' approach has been 
studied here and in \cite{m2c, hybrid, Cheng:2008hk}. As summarized 
below, adding the dilepton edge information improves the two-chain 
mass determinations obtained following the basic procedure developed 
in this paper. 
 
In our procedure, we have applied a set of bias reduction methods. In  
particular, to reduce the number of combinations, we have utilized the 
fact that correct combinations can pair with relatively more 
events than wrong combinations. Alternatively, one may try to 
reduce the number of wrong combinations before doing the event 
reconstruction. For example, one could try to group objects into two 
hemispheres \cite{hemisphere} and assume that the objects in the same 
hemisphere come from the same decay chain. However, this only works 
well when the initial particles are substantially boosted, while the squarks in 
our case are produced mostly close to the threshold without large 
boosts. For small boosts, the quark and $\cntwo$ from squark decay 
actually belong to two opposite hemispheres instead of the same one. The 
directions of the subsequent $\cntwo$ and $\slep$ decay products are 
even more random. We have applied the hemisphere method on a set of 
ideal events from squark pair production, each containing 2 quarks and 
4 leptons according to the decay chain in 
Fig.~\ref{fig:chain_decay}. Even without any complications from extra 
jets, experimental smearing and so forth, only about 12\% of the events have 
the decay chains correctly identified (this does not account for the 
ambiguity of the two leptons in the same decay chain).   
 
One could explore the effectiveness of imposing a cut which accepts 
only events with substantial thrust or small circularity before 
separating each event into two hemispheres.  It is not clear to us 
that the gain from decreasing the combinatorics problem would outweigh 
the reduced statistics associated with the fact that such a cut would 
remove a large fraction of the available events. 
 
However, we {\it have} shown that when a two-particle mass edge (such as the 
dilepton mass edge in the examples we have considered) can be 
identified, it is very useful to impose a cut whereby only solutions 
that give a dilepton mass within roughly $\pm 20\gev$ of the edge 
location are retained.  By applying this cut before proceeding with 
the rest of our analysis procedure the mass peak biases are 
essentially eliminated and the errors on the central mass values are 
very similar. In addition, we have shown that this cut is 
capable of essentially eliminating contamination of the mass 
determinations for the SUSY signal of interest by events coming from 
some other SUSY signal that does not share the same dilepton edge 
location.  Presumably any other recognizable kinematic edge could be 
exploited in similar fashion, but dilepton mass edges will typically 
be least impacted by detector momentum smearing.

In conclusion, we have proposed a kinematic technique for mass 
determination in events with two invisible dark matter particles. The 
technique seeks constraints on the mass space from measured momenta. 
In Sec.~\ref{sec:counting}, we have given general constraint counting 
and discuss the corresponding strategies for both the single decay 
chain case and the double decay chain case. In the former, one only 
uses the information from one of the two decay chains in each event; 
in the latter, one uses information associated with both decay chains. 
The constraints include the mass-shell constraints for the dark matter 
particle as well as all intermediate particles. In the double decay 
chain case, we obtain extra constraints from the measured missing 
transverse momenta. In both cases, more constraints are available when 
the decay chain is longer. In certain instances (this includes the 
single decay chain case with 3 visible particles per chain and the 
double identical decay chain case with 2 visible particles per chain), 
we obtain discrete solutions for the missing momenta by using trial 
masses for the unknown particles. Requiring that we obtain physical 
solutions for the momenta, the consistent mass region becomes more 
restricted when more events are included. The actual masses are 
obtained by studying the kinematic distributions of the consistent 
region. This is the strategy we adopted in Refs.~\cite{Cheng:2007xv, Cheng:2008hk}. 
When the decay chains present in each event are longer, it is possible 
to obtain discrete solutions for the momenta (and therefore the 
masses) by combining the constraints from a few different events, {\it 
  without} assuming any trial masses. This occurs when there are 4 or 
more visible particles per decay chain for the single chain case, or 3 
or more visible particles per decay chain for the double chain case. 
In this article, we have focused on the double identical chain case 
with the number of visible particles per chain fixed to 3. 
 
In our case, the constraints can be solved for discrete solutions of 
the unknown masses when two events are combined. However, because the 
system of equations contains quadratic equations, wrong solutions are 
introduced. Nonetheless, if the visible momenta could be measured 
without errors, it would take only three events to obtain the correct 
masses. This is because the wrong solutions would be different for the 
three different possible event pairings whereas the correct solution 
remains the same for every event pair. This remains true even when 
wrong combinations are included. However, in practice the non-zero 
experimental resolutions imply that the correct solution distribution 
becomes smeared which then overlaps with the distribution coming from 
wrong solutions, wrong combinations and background events. Despite 
this, we have shown that when two of the visible particles in a decay 
chain are leptons, we obtain good precision ($\sim$ a few GeV) if a 
few hundred events are available. We have developed methods to reduce 
the number of wrong combinations and backgrounds. The resulting mass 
solution distributions are clearly peaked around the input masses with 
small systematic errors which can be eliminated either by comparing 
with Monte Carlo distributions around the estimated masses or by 
imposing an initial dilepton edge mass cut on the accepted solutions. 
An important assumption for doing this comparison is that the 
distributions are only sensitive to the masses instead of the 
underlying theories. We have shown that this is indeed the case by 
comparing two distinct theories with different spin structure, MSSM 
and UED. We have set the spectra of the two models to be the same and 
examined the mass distributions, which show little difference. 
Correspondingly, the systematic errors introduced by model dependence 
are much smaller than the statistical errors.

Finally, we comment on possible improvements of our method. Given the
precision of the purely kinematic results and availability of spin
determination techniques \cite{spin}
it would certainly be possible to figure out the
underlying theory/spins and then apply model-dependent techniques to
refine the mass determinations. For example, one could adopt a
likelihood method similar to the one used in the top mass measurement
with dilepton events at Tevatron \cite{topmass}. In this method, a
probability density, as a function of all unknown masses, is defined
for each event by convoluting matrix elements and detector resolution
functions. One then obtains the joint probability by taking the product of
the probability densities from all events. The best-fit masses are
given by the values that maximize the joint probability. Compared with
top quark mass measurement, the event topology considered in this
paper is more complicated since that more final state particles are
involved. One also needs to scan a four dimensional instead of one
dimensional mass space to minimize the probability function. Therefore,
another benefit of the purely kinematic method is that it
significantly simplify the computation by reducing the candidate mass
space to a very small region.

We could also consider simplified likelihood methods using the detector
resolution functions, but without the knowledge of the matrix
element. One such method is based on the same constraints,
Eqs.~(\ref{mass_shell}) through (\ref{misspt}), which can be viewed as
a generalization of the method discussed in this paper: for each
event, we can eliminate the 4-momenta of the two missing particles by
using Eqs.~(\ref{mass_shell})-(\ref{misspt}). We are then left with two
equations in the form 
\begin{eqnarray}
f(m_N, m_X, m_Y, m_Z; p_{vis}) &=& 0,\\
g(m_N, m_X, m_Y, m_Z; p_{vis}) &=& 0,
\end{eqnarray}
where $f$ and $g$ are functions of the masses and all visible momenta,
$p_{vis}$. If we require that 
the equalities hold exactly, then each event defines a 2-dimensional
surface in a 4-dimensional mass space. The 2-dimensional surfaces of
two different events intersect at discrete points, which are nothing
but the mass solutions. This is equivalent to saying
that in the mass space, for each event we assign a non-zero
probability density for points on
the surface and a zero probability density for points off the surface. Then by
combining two events, the joint probability is non-zero only at
discrete points. Obviously, a more sophisticated method is to assign a
maximum probability density for points on the surface, and smaller but non-zero
probability densities for points away from the surface. This can be
done by calculating the $\chi^2$ distribution for each point in the
mass space,
\begin{equation}
\chi^2=\frac{f^2}{\sigma_f^2}+\frac{g^2}{\sigma_g^2},
\end{equation}  
where 
\begin{equation}
\sigma_f^2\equiv\sum_{p_{vis}}\left(\frac{\partial f}{\partial
  p_{vis}}\sigma_{p_{vis}}\right)^2,\label{eq:error}
\end{equation}
and a similar formula holds for $\sigma_g$. Here, we have assumed
Gaussian distributions for the invisible momenta $p_{vis}$ with errors
given by $\sigma_{p_{vis}}$, and Eq.~(\ref{eq:error}) can be
complicated by correlations among the visible momenta. We then 
sum the $\chi^2$ over all events and obtain the
masses when the total $\chi^2$ is minimized. In practice, $f$ and $g$
are very complicated functions of the visible momenta, which may
result in large roundoff errors. As in the matrix element method, one
also needs to efficiently minimize the $\chi^2$ over a 4-dimensional
space. Therefore, it deserves further studies to determine whether
this approach can improve the precision of the mass measurement over
the simple method discussed in this paper.

In closing, we note that the program for determining the $m_{Z,Y,Z,N}$ 
solutions as a function of input visible momenta for two-chain events 
is available at 
http://particle.physics.ucdavis.edu/hefti/projects/ (choose {\it
  WIMPMASS}). Stand-alone programs that implement the methods of
Refs.~\cite{Cheng:2007xv, Cheng:2008hk} are also available at this
same website. 
 
\acknowledgments  
This work was supported in part by the U.S.~Department of Energy grant 
No.~DE-FG02-91ER40674.  
\appendix 
 
\section{Solving the constraint equations} 
\label{app:solve} 
In this Appendix we describe in detail our 
procedure to solve the system of the kinematic constraint equations 
\begin{eqnarray} 
&&p_1^2=p_2^2=q_1^2=q_2^2,\\ 
&&2p_1\cdot p_3+p_3^2=2p_2\cdot p_4+p_4^2=2q_1\cdot q_3+q_3^2=2q_2\cdot q_4+q_4^2,\\ 
&&2(p_1+p_3)\cdot p_5 +p_5^2=2(p_2+p_4)\cdot p_6+p_6^2=\nonumber\\ 
&&\    \ =2(q_1+q_3)\cdot q_5+q_5^2=2(q_2+q_4)\cdot q_6+q_6^2,\\ 
&&2(p_1+p_3+p_5)\cdot p_7 +p_7^2=2(p_2+p_4+p_6)\cdot p_8+p_8^2=\nonumber\\ 
&&\    \ =2(q_1+q_3+q_5)\cdot q_7+q_7^2=2(q_2+q_4+q_6)\cdot q_8+q_8^2,\\ 
&&p_1^x+p_2^x=p_{miss}^x,\  \ p_1^y+p_2^y=p_{miss}^y,\\ 
&&q_1^x+q_2^x=q_{miss}^x,\  \ q_1^y+q_2^y=q_{miss}^y\,. 
\end{eqnarray}  
As mentioned earlier, it is straightforward to numerically solve the 
above equations using commercial software such as Mathematica, but the 
speed is intolerably low. Instead, we solve the equation system using a 
programming language such as $\CC$. The idea is to reduce the system 
to a univariate polynomial equation whose coefficients are (fixed) 
functions of the original visible momenta. Note that it is convenient 
to obtain the coefficient functions with the assitance of 
Mathematica. After 
that, the functions are hard coded in the $\CC$ program and Mathematica 
is no longer needed. The univariate equation can then be solved 
numerically using any available polynomial solver. The Mathematica 
notebook and the $\CC$ code are available from Ref.~\cite{code} or any of 
the authors.  The key Mathematica operations are also described in 
Appendix \ref{app:mathematica} for reference. The method can be easily 
generalized to solve other polynomial equations efficiently. 
   
It is straightforward to eliminate 13 variables using the 13 linear 
equations and obtain 3 quadratic equations with 3 variables. 
Generically, 3 quadratic equations can be written as 
\begin{eqnarray} 
 z^2+a_8 z y+a_7 z x+a_6 z+ a_5 y^2+a_4 yx+a_3 y+a_2 x^2+a_1 x+a_0&=&0,\label{eq1}\\ 
z y+b_7 z x+b_6 z+ b_5 y^2+b_4yx+b_3 y+b_2 x^2+b_1 x+b_0&=&0,\label{eq2}\\ 
z x+c_6 z+c_5 y^2+c_4 yx+c_3y+c_2x^2+c_1x+c_0 &=&0.\label{eq3} 
\end{eqnarray} 
where $x,y,z$ are variables, and $a_i,b_i,c_i$ are coefficients as 
functions of the original visible momenta. We 
have ordered the left-hand terms lexicographically in the order 
$z>y>x$. We will eliminate variables also in this order and eventually 
obtain a univariate equation in $x$. In our implementation, we choose 
$x,y,z$ to be $p_1^0,p_2^0,q_1^0$ or some permutation thereof.  In fact, we 
solve several times for several different permutations just to make 
sure we do not miss any solutions. 
 
First, by calculating $\mbox{(\ref{eq1})}\times 
y-\mbox{(\ref{eq2})}\times z$, we cancel the term $z^2 y$ and obtain a 
polynomial equation with leading term $-b_7xz^2$. We repeatedly use 
(\ref{eq1}), (\ref{eq2}), (\ref{eq3}) to reduce the polynomial, i.e., 
to eliminate the leading term of the polynomial. For example, we 
eliminate the $-b_7xz^2$ term by subtracting 
$-b_7\times\mbox{(\ref{eq3})}$. The next leading term is $\propto 
z^2$, which again can be eliminated by subtracting from it 
Eq.~(\ref{eq1}) with appropriate coefficient.  Repeating this 
procedure until it cannot be reduced further, we obtain an equation in 
the form 
\begin{equation} 
z+y^3+y^2 x+y ^2+y x^2+y x+y+x^3+x^2+x+1=0,\label{eq4} 
\end{equation} 
where we have omitted all coefficients. Similarly, reducing 
$\mbox{(\ref{eq1})}\times x-\mbox{(\ref{eq2})}\times z$ , we obtain 
another equation also in the form of (\ref{eq4}).  Canceling the 
leading term, we eliminate the variable $z$ and obtain a cubic 
equation in $(y,x)$.  Reducing $\mbox{(\ref{eq2})}\times x - 
\mbox{(\ref{eq3})}\times y$, we obtain another cubic equation in 
$(y,x)$.

We are then left with 2 cubic equations in 2 variables, 
\begin{eqnarray} 
y^3+d_8 y^2x+d_7y^2+d_6y x^2+d_5yx+d_4y+d_3x^3+d_2x^2+d_1x+d_0=0,\nonumber\\ 
y^2x+e_7y^2+e_6y x^2+e_5yx+e_4y+e_3x^3+e_2x^2+e_1x+e_0=0,\label{cubic} 
\end{eqnarray} 
where the coefficients $d_i$ and $e_i$ are derived from $a_i$, $b_i$ 
and $c_i$ in Eq.~(\ref{eq1}) (\ref{eq2}) (\ref{eq3}). This system can 
be solved using the following method. The resultant \cite{resultant} 
of two univariate polynomials is defined as follows.  Given a 
polynomial (note that the $a_i$ and $b_i$ below are not those given in 
Eq.~(\ref{eq1}) and (\ref{eq2})) 
\beq 
P(x)=a_nx^n+a_{n-1}x^{n-1}+...+a_1x+a_0\,, 
\eeq 
of degree $n$ with roots $\alpha_i, \quad i=1, ..., n$ and a polynomial 
\beq 
Q(x)=b_mx^m+b_{m-1}x^{m-1}+...+b_1x+b_0          
\eeq 
of degree $m$ with roots $\beta_j, \quad j=1, ..., m$, the resultant 
$\rho(P,Q)$, also denoted $R(P,Q)$ and also called the eliminant, is 
defined by 
\beq  
\rho(P,Q)=a_n^mb_m^n\prod_{i=1}^n\prod_{j=1}^m(\alpha_i-\beta_j)        
\label{resultant} 
\eeq 
 
The resultant is also given by the determinant of the corresponding 
Sylvester matrix \cite{sylvester}.  
The Sylvester matrix associated to polynomials $P$ and $Q$ is the 
$(n+m)\times(n+m)$ matrix obtained as follows: 
\ben 
\item 
    the first row is: 
\beq 
    \begin{pmatrix} a_n & a_{n-1} & \cdots & a_1 & a_0 & 0 & \cdots & 0 \end{pmatrix}. 
\eeq 
\item 
     the second row is the first row, shifted one column to the right; the first element of the row is zero. 
\item 
 the following $(m-2)$ rows are obtained the same way, still filling 
 the first column with a zero. 
\item 
    the $(m+1)$-th row is: 
\beq 
    \begin{pmatrix} b_m & b_{m-1} & \cdots & b_1 & b_0 & 0 & \cdots & 0 \end{pmatrix}. 
\eeq 
\item 
     the following rows are obtained the same way as before. 
\een 
Taking the left-hand side polynomials of Eq.~(\ref{cubic}) as $P$ and 
$Q$ respectively, we have $n=3$ and $m=2$ and the Sylvester matrix in 
this case is 
\begin{equation} 
S=\left ( 
\begin{array}{ccccc}a_3&a_2&a_1&a_0&0\\0&a_3&a_2&a_1&a_0\\b_2&b_1&b_0&0&0\\0&b_2&b_1&b_0&0\\0&0&b_2&b_1&b_0\end{array}\right 
 ),  
\end{equation} 
where  
\begin{eqnarray} 
&a_3=1,\ \ a_2=d_8x+d_7,\ \ a_1=d_6x^2+d_5x+d_4,\ \ a_0=d_3x^3+d_2x^2+d_1x+d_0,\nonumber\\ 
&b_2=x+e_7,\ \ b_1=e_6x^2+e_5x+e_4,\ \ b_0=e_3x^3+e_2x^2+e_1x+e_0. 
\end{eqnarray} 
Therefore, $\rho(P,Q)=\mbox{det}S$ is a 9th order polynomial in $x$. 
By Eq.~(\ref{resultant}), if $\rho(P,Q)=0$, the two equations in 
(\ref{cubic}) have at least one common root. In other words, for each 
root $x_i$ of $\mbox{det}S=0$, we can find a $y_i$ such that 
$(x_i,y_i)$ is a solution of Eqs.~(\ref{cubic}). Thus, the problem has 
been reduced to the solution of a 9th order polynomial equation with 
real coefficients. We solve it numerically using algorithm TOMS/493 
\cite{toms493}. Note that one of the roots is fake.  It can easily be 
identified and eliminated by substituting back each of the solutions into 
the original equations to identify the root that is not actually a 
solution of the original equations.  
 
\section{Mathematica file} 
\label{app:mathematica} 
Many of the operations described in Appendix \ref{app:solve} can be 
conveniently done in Mathematica, which we describe below. 
 
First, suppose we have obtained the equations (\ref{eq1}), 
(\ref{eq2}) and (\ref{eq3}). As mentioned, the coefficients $a_i$, $b_i$ 
and $c_i$ are functions of the visible momenta. These functions are 
usually complicated, therefore it is always desirable 
to use intermediate parameters such as $a_i$, $b_i$ 
and $c_i$, and normalize the coefficients of the leading terms to 1. To obtain 
Eqs.~(\ref{cubic}), we use the Mathematica function PolynomialReduce: 
{\small 
\begin{eqnarray*} 
&&P1 = a0 + a1\, x + a2\, x^2 + a3\, y + a4\, y\, x + a5\, y^2 + a6\, z + a7\, z\, x + a8\, z\, y +  z^2;\\ 
&&P2 = b0 + b1\, x + b2\, x^2 + b3\, y + b4\, y\, x + b5\, y^2 + b6\, z + b7\, z\, x + z\, y;\\ 
&&P3 = c0 + c1\, x + c2\, x^2 + c3\, y + c4\, y\, x + c5\, y^2 + c6\, z + z\, x;\\ 
&&P4 = Expand[ 
   PolynomialReduce[P1\,y - P2\,z,\, \{P1,\, P2,\, P3\},\, \{z,\, y,\, x\}][[2]]];\\ 
&&P5 = Expand[ 
   PolynomialReduce[P1\,x - P3\,z,\, \{P1,\, P2,\, P3\},\, \{z,\, y,\, x\}][[2]]];\\ 
&&P6 = Expand[ 
   PolynomialReduce[P2\,x - P3\,y,\, \{P1,\, P2,\, P3\},\, \{z,\, y,\, x\}][[2]]]; 
\end{eqnarray*} 
} 
where we obtain $P4$ and $P5$ as two polynomials in the form of 
Eq.~(\ref{eq4}) and $P6$ a cubic polynomial in $x$ and $y$. Then it is 
straightforward to cancel the variable $z$ and cast the remaining two 
polynomials in the form of Eq.~(\ref{cubic}). The resultant is obtained 
from 
{\small 
\begin{eqnarray*} 
&&P7 = d0 + d1\, x + d2\, x^2 + d3\, x^3 + d4\, y + d5\, y\, x + d6\, y\, x^2 + d7\, y^2 + 
    d8\, y^2\, x + y^3;\\ 
&&P8 = e0 + e1\, x + e2\, x^2 + e3\, x^3 + e4\, y + e5\, y\, x + e6\, y\, x^2 + e7\, y^2 + 
    y^2\, x;\\ 
&&resultant = Resultant[P7,\, P8,\, y]; 
\end{eqnarray*}} 
where $resultant$ is the 9th order polynomial in $x$ we are 
seeking. After obtaining the roots of the resultant numerically, $y$ and 
$z$ can be uniquely determined as follows: defining the polynomial $P9$ by 
{\small 
\begin{equation*} 
P9 = Expand[PolynomialReduce[P7\,x - P8\,y,\, \{P7,\, P8\},\, \{y,\, x\}][[2]]]; 
\end{equation*} 
} 
we see that $P9$ can be written in the form  
{\small 
\begin{equation*} 
P9 = f0 + f1\, x + f2\, x^2 + f3\, x^3 + f4\, x^4 + f5\, y + f6\,  y\, x +  
   f7\, y\, x^2 + f8\, y\, x^3 + y^2; 
\end{equation*} 
} 
It turns out that the polynomial $P10$ defined below is linear in $y$: 
{\small 
\begin{equation*} 
P10 = Expand[ 
   PolynomialReduce[P9\,x - P8,\, \{P8,\, P9\},\, \{y,\, x\}][[2]]]; 
\end{equation*} 
} 
We then obtain $y$ from the equation $P10=0$ by substituting in the 
solutions for $x$. Once this is done, we can obtain $z$ from the equation $P4=0$ 
since it is linear in $z$. One of the 9 solutions is a fake solution 
to the original equation system, which can be easily identified by 
finding the solution which does not actually solve the system of 
equations when substituted back into the system of equations.

\end{document}